\documentclass{aastex63}
\usepackage{soul}
\usepackage[normalem]{ulem}

\newcommand{\vishal}[1]{#1}
\newcommand{\newref}[1]{#1}
\newcommand{\thirdround}[1]{#1}
\newcommand{\psm}{\texttt{PSM}}

\received{}
\revised{}
\accepted{}

\submitjournal{ApJ}
\shorttitle{On the Impulsive Heating of Quiet Solar Corona}
\shortauthors{Upendran and Tripathi}
\begin{document}
\title{On the Impulsive Heating of Quiet Solar Corona}
\author[0000-0002-9253-6093]{Vishal Upendran}
\author[0000-0003-1689-6254]{Durgesh Tripathi}
\affiliation{Inter University Centre for Astronomy and Astrophysics, Pune, India - 411007}

\begin{abstract}
The solar corona consists of a million-degree Kelvin plasma. A complete understanding of this phenomenon demands the study of Quiet Sun (QS) regions. In this work, we study QS regions in the 171~{\AA}, 193~{\AA} and 211~{\AA} passbands of the Atmospheric Imaging Assembly (AIA) on board the Solar Dynamics Observatory (SDO), by combining the empirical impulsive heating forward model of \cite{PSModel} with a machine-learning inversion model that allows uncertainty quantification. We find \thirdround{that there are  $\approx 2-3$ impulsive events per min}, with a lifetime of about $10-20$ min. Moreover, for all the three passbands, the distribution of power law slope $\alpha$ peaks above 2. Our exploration of correlations among the \thirdround{frequency of impulsive events and their timescales} and peak energy suggests that conduction losses dominate over radiative cooling losses. All these finding suggest that impulsive heating is a viable heating mechanism in QS corona.  
\end{abstract}
\keywords{Sun, Corona, Atmosphere}
\section{Introduction}\label{sec:intro}
Full disk images of the Sun taken in extreme ultraviolet (EUV) and X-rays, consist of three features: Active Regions (ARs),  Coronal holes (CHs), and Quiet Sun (QS). The ARs are the brightest, CHs appear darkest, whereas QS regions appear like a background. Note that while the number of ARs is a function of solar activity cycle, QS is present irrespective of the activity phase. There is a lot of literature studying the coronal heating in the ARs \citep[see][for a review]{fabioreale_arloops} -- however, studies of coronal heating in QS and CHs are sparse \cite[see e.g.][]{TriNS_2020}. To address the physics of coronal heating in general, a comprehensive study of the heating in QS regions is of paramount importance\citep{ASCHWANDEN2014235}.

One of the most popular mechanisms to heat the corona is via impulsive events , viz. Nanoflares \citep[see e.g.][]{parker1988nanoflares}. Impulsive events are transients generated through the dissipation of magnetic stresses or waves \citep[see e.g.,][]{AntolinShibata}. Solar atmosphere presents us with plethora of impulsive events at various energies and time scale, viz. flares~\citep{benz2006, TriBC_2004}, microflares~\citep{hannahmicroflares, 1983xrays, srividya, chifor_2006, chifor}, active region transient brightenings~\citep{testaARtransient, GupST_2018, Tri_2021, nivedgerrypowerlaw, rajhans2021}, transition region blinkers~\citep{harrison}, UV bombs~\citep{hardipeter_heating,  GirjeshTripathi}, Ellerman bomb~\citep{ellerman1917, pariatEB, isobetripathiarchontis} and other activities such as jets~\citep{MulTDM_2016, Raouafi2016}. It is also observed that the properties of loops in ARs are well described by the impulsive heating scenario~\citep[][ and references therein]{Ghosh_2017, Tripathi2009, warren2008, GirjeshTripathi, winebarger2013}. \vishal{Such a scenario has also been studied independently in ARs using a variety to techniques like Time lag analysis~\citep{viall2012, viall2013,viall2015,viall2016,Viall2017}, Differential Emission Measure (DEM) and Doppler shifts analysis \citep[see e.g.,][]{TriMK_2010, tripathi2011, TriMK_2012, winebarger2011, warren2012, WinTMD_2013, SubTKM_2014, delGT_2015}, hydrodynamic modelling \citep[see e.g.][]{bradshaw2012,reep2013,cargill2015,barnes2016}, Magneto hydrodynamic modelling \citep[see e.g.][]{rappazzo2015,rappazzo2017,knizhnik2018,knishnik2019,knizhnik2020} and empirical models~\citep[see eg.][]{jess2019,Jess_2014_chromosphere}. Thus,  it is natural to assume that the coronal heating in QS may also be governed by impulsive heating. However, since the QS regions have a very diffuse structure, it is not possible to count individual events and understand the energetics of these events in the QS.}

To maintain the corona at a million degree, the frequency distribution of impulsive events must follow a power law distribution in energy -- i.e., $\frac{dN}{dW}\propto W^{-\alpha},$ with $\alpha > 2$ \citep[see][]{hudson1991solar}. Observations do show that impulsive events in the corona follow a power-law distribution -- however, there are a range of $\alpha$ values reported in the literature~\citep[see, for example, Fig. 6.14 in][]{Aschwanden2019}. 

One of the most significant caveats in these results is due to the assumption that the each detected event is a single entity. But, we know that small-scale impulsive events may happen at a sub-resolution scale. Hence, individual brightenings may not be single entity but may consist of many tiny events. This may lead to the under counting of events particularly at lower energies. Moreover, the observations of QS radiance in UV and EUV, both in space and time, show log-normal distribution \citep{ps_2001}. Thus, the the QS radiance might be generated due to some form of a Markovian process~\citep{PSModel,gorobets2016markov}.

To mitigate the the above mentioned issues, \citet{PSModel} proposed an empirical model for the heating of the QS corona. Hereafter, we call it the \texttt{Pauluhn and Solanki Model (PSM)}. In brief, \texttt{PSM} approximates the response of the plasma to a unit heating event as an exponential rise and fall of intensity. The amplitude of the heating event is sampled from a power-law distribution, with the frequency of occurrence of the events being kept as a free parameter. The resultant light curve is then a combination of a multitude of these events. Hence this addresses the sub-pixel resolution scale of these structures. Moreover, the resulting intensities are also shown to be log-normal, mimicking the observations.~\vishal{Finally, the observed power-law distribution of energetic events is also incorporated in this model, thereby enabling us to understand the viability of impulsive events maintaining the observed intensity in a given light curve.} Thus, \texttt{PSM} may provide an excellent proxy for the generation of the QS coronal intensities.

\cite{PSModel} generate the parameters for different observations by comparing similarity of intensity distribution and the Global Morl\'et wavelet power~\citep{torrence1998practical} of simulated light curves with those of observed light curves. The comparison is qualitative -- a sufficient good match in distribution, and the frequency location of peaks in the power spectrum were taken to represent a good match between the observation and simulation. Although the comparison had a sound basis, it was done by eye, and needs to be on a more quantitative foundation.


\vishal{The problem of obtaining parameters from the observed light curve thus becomes an inversion problem. } In general, the inversion approaches depend primarily on generating important ``features'' from the light curves which then have a one to one mapping with the parameter set of \texttt{PSM}. This is essentially performed qualitatively by \cite{PSModel}. Since it is quite non trivial to objectively pick out features for inversion, one trains an inversion model to pick out abstract features, and perform the inversion. In this case, an optimization principle guides the mapping from light curves to parameter set developed by the inversion model.

\citet{safari1, safari2} employed this method by using a Probabilistic Neural Network (PNN). Under this scheme, every simulated light curve is classified, where each class has a unique combination of free parameters. The PNN is trained on the full set of simulated light curves. Finally, the observed light curves are fed into the PNN, which then assigns each of them to one of the learned classes. For their study, \cite{safari1} used $\approx$ 10,000 light curves (at max). These light curves corresponded to CHs, QS, and ARs on the Sun obtained from the Atmospheric Imaging Assembly \citep[AIA;][]{Lemen2012} on board Solar Dynamics Observatory \citep[SDO;][]{SDO} and data from the Extreme UltraViolet Imager (EUVI) on board STEREO~\citep{kaiser2008stereo}. On average, they obtained $\alpha > 2$ for all the regions.

This method is a great start to a tricky stochastic inversion problem. But, such an approach must be well-validated by an exhaustive testing set. Moreover, the classification of light curves imposes a discretisation constraint on the parameter set, which depends on the grid resolution of the simulation. Thus, any assigned class to a given light curve may change on improving the grid resolution. In other words, we do not know the confidence level of the PNN for each inversion. 

We further emphasize that \cite{safari1} used poor resolution of AIA 171~{\AA} data ($90$s cadence and $2$\arcsec spatial resolution), and further binned the data spatially by $3\times3$ and $5\times5$ window. Such a binning will average out the small scale events, and thus, will not allow the use of full potential of AIA observations. Here we use the full spatial and temporal resolution observations taken using AIA and perform a regression on the target parameter set using a Convolutional Neural Network \citep[CNN;][]{lecun2015deep}, that circumvents the potential caveats in the work of \citet{safari1, safari2} as detailed above.

A CNN is used to preserve information of the time scales of features in the light curve along with the distribution of intensity values. A CNN may be thought of as a set of `kernels', which perform convolution on the input and return a convolved output. The kernel size can be interpreted as the scale over which information is summarized. Several such kernels are operated on a given input, and these outputs are passed through a non-linear function called the `Activation function'. Successive kernels are thus sensitive to local scales of the corresponding input, but that input itself may be an extremely non-linear transformation of the original light curve. 

Furthermore, we do not use a simple CNN for the inversion. Instead, we use an Inception module, which improves its performance \citep[see,][]{szegedy2015going}. The Inception module provides kernels with multiple sizes at the same level and makes is possible to perform multi-scale analysis at each level. By employing a CNN to capture the scales of variation present in the data, we circumvent grid resolution issues by considering the target as a regression problem. We can effectively interpolate between simulated grid points while performing the inversion. With the CNN, we also obtain associated inversion uncertainties by perturbing the network.

The rest of the paper is arranged as follows. We describe our datasets and associated noise in \S\ref{sec:data} \& \S\ref{sec:noise}, respectively. The modelling and validation of the network is discusses in \S\ref{sec:methods}. The results are presented in \S\ref{sec:results}, followed by a summary and conclusions in Sec.~\ref{sec:discussion}. 

\section{Observations and Data}\label{sec:data}

For this work, we have used the observations recorded by AIA onboard SDO. AIA observes the Sun's atmosphere in UV and EUV bands using eight different passbands sensitive to plasma at different temperatures \citep{OdwDM_2010, BoeEL_2012}. Here we have used the data taken using 171~{\AA}, 193~{\AA} and 211~{\AA} passbands. These images are taken with a pixel size of $\sim$0.6{\arcsec} and an approximate time cadence of 12~s. We have chosen these particular passbands since the count rates in these passbands for QS is large, when compared to others~\citep[see Table 2 in][]{ODwyer}.

\begin{figure*}[htpb!]
\includegraphics[width=0.6\linewidth]{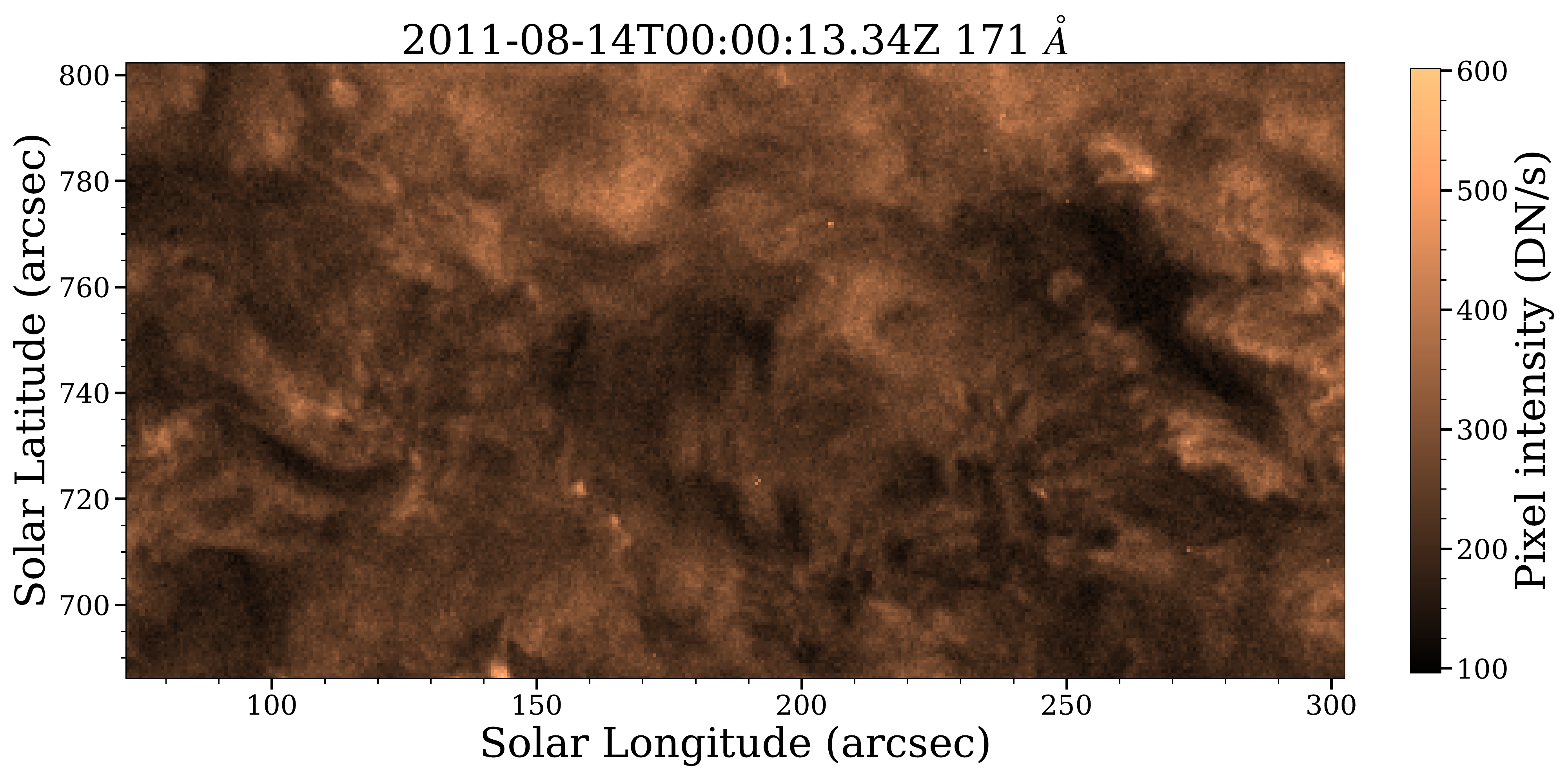}
\includegraphics[width=0.3\linewidth]{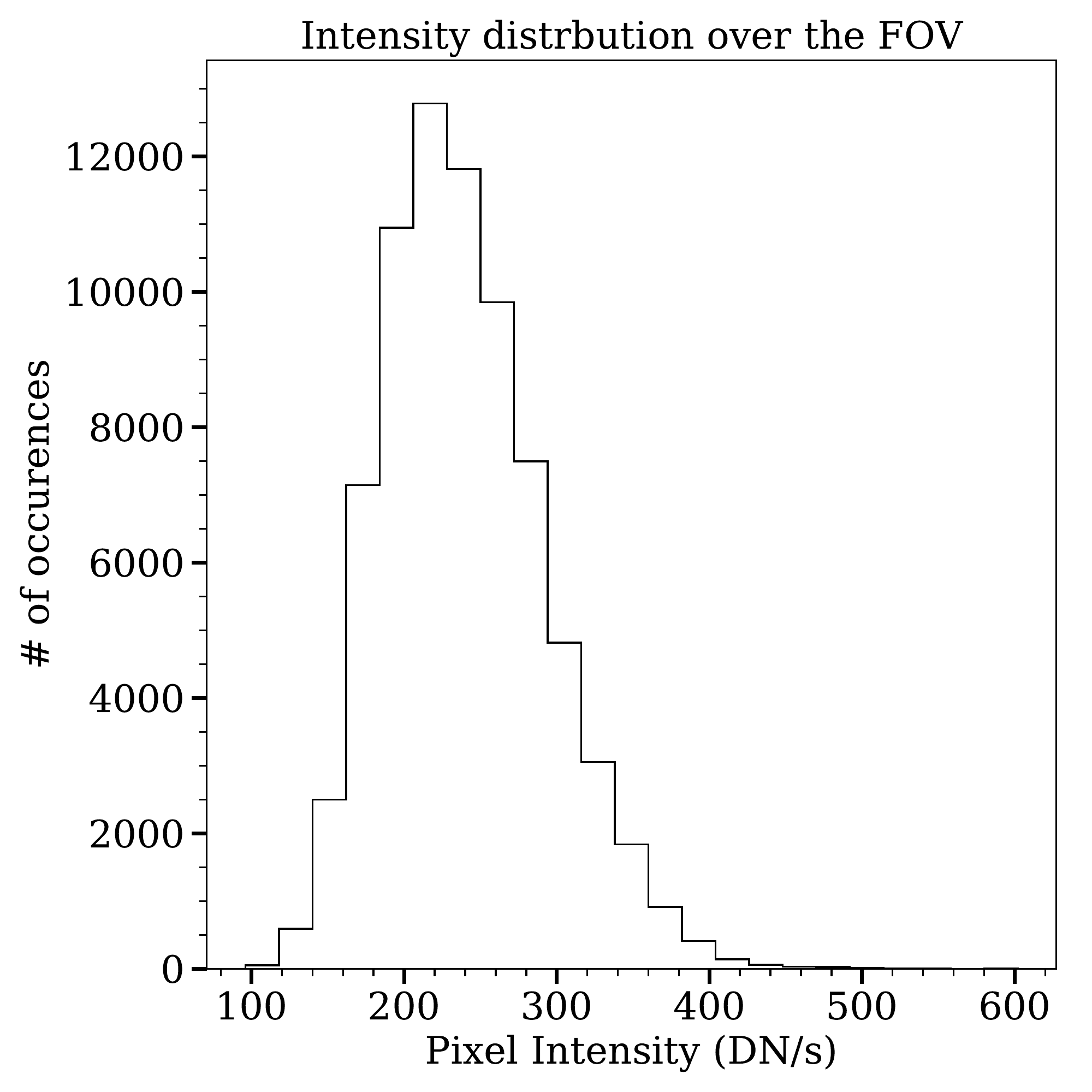}
\caption{171~{\AA} image of quiet Sun corresponding to DS1, and the corresponding \vishal{histogram of } intensity.} \label{fig:FOV1}
\end{figure*}

\begin{figure*}[htpb!]
\includegraphics[width=0.52\linewidth]{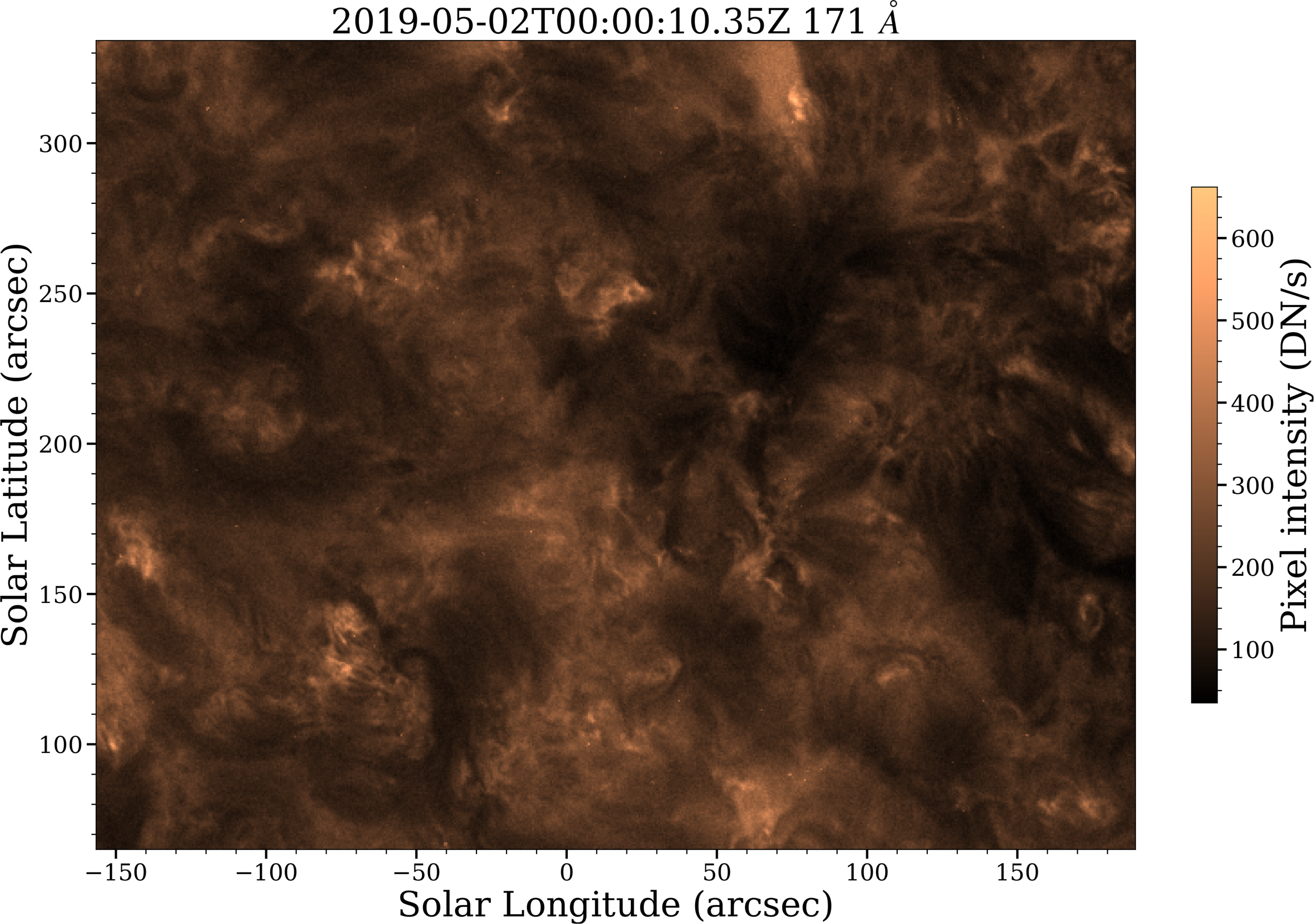}
\includegraphics[width=0.38\linewidth]{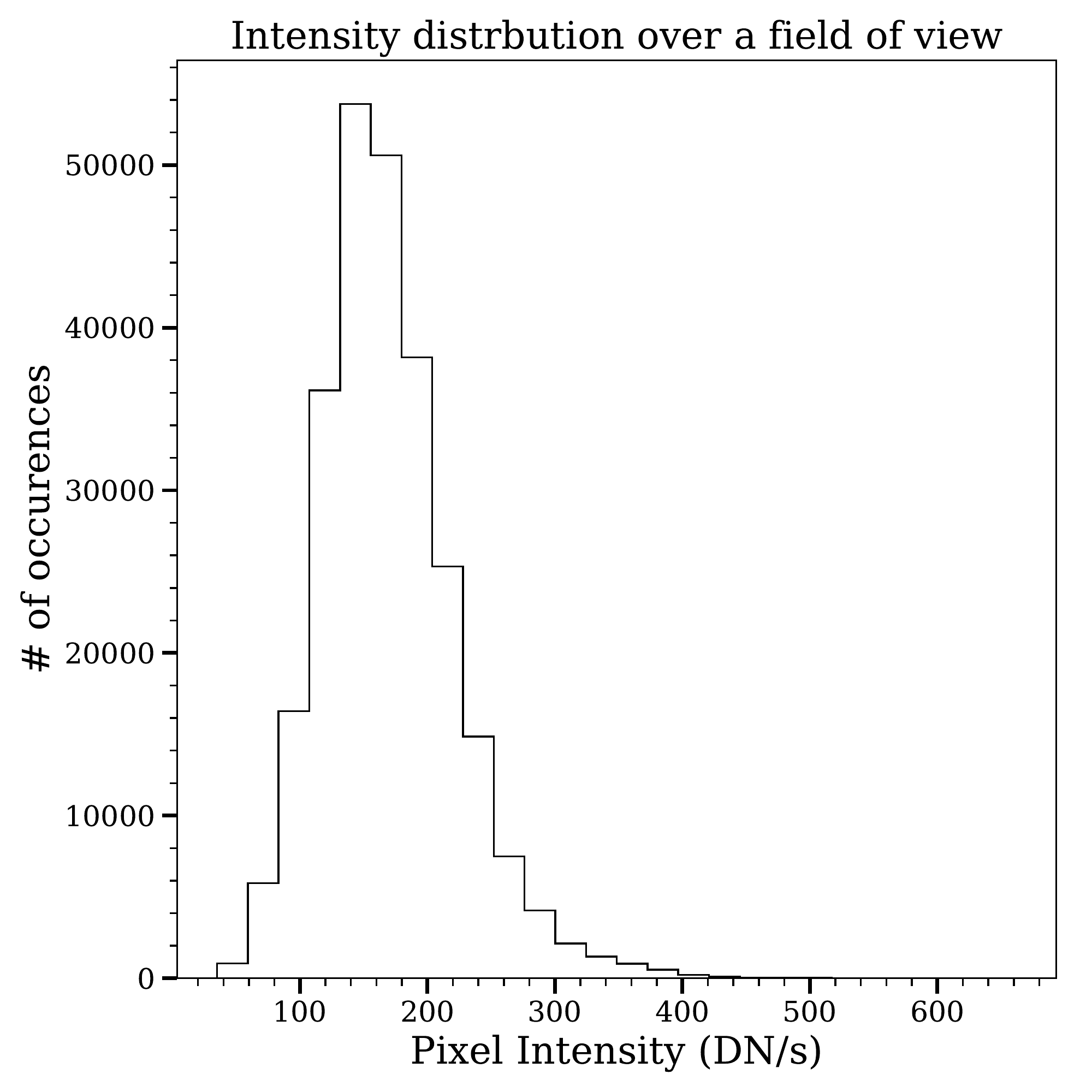}
\caption{Same as Fig.~\ref{fig:FOV1} but for DS2.} \label{fig:FOV2}
\end{figure*}

Monitoring the EUV images on \texttt{Solar Monitor} \footnote{https://www.solarmonitor.org/}, we have identified QS patches during 2011 and 2019, where no activity was observed. Details of the two data sets (DS1 \& DS2) are given in Table.~\ref{tab:2011data}. In total, we have obtained eight continuous hours of data for each set, which corresponds to 2400 time steps. All the images are aligned to the first image and are exposure time normalized. The full FOV (single snapshot) for DS1 and DS2 as observed in 171~{\AA} is displayed in the left panel of Figs.~\ref{fig:FOV1} \& \ref{fig:FOV2} with the corresponding spatial distribution of intensities in the right panels.

\begin{deluxetable*}{|c|c|c|}
\tablecaption{Dataset details \label{tab:2011data}}
\tablewidth{0pt}
\tablehead{\colhead{Identifier} & \colhead{DS 1} & \colhead{DS 2}}
\startdata
Start time & 2011-08-14 T00:00:00 & 2019-05-02 T00:00:00 \\
End time & 2011-08-14 T08:00:00 & 2019-05-02 T08:00:00 \\
Reference time & 2011-08-14 T00:00:00 & 2019-05-02 T00:00:00\\
Xcen,Ycen  & \thirdround{192\arcsec, 749\arcsec} & \thirdround{19.0\arcsec, 211.5\arcsec}\\
FOVx, FOVy & \thirdround{230\arcsec, 116\arcsec}  & \thirdround{346.0\arcsec, 269.0\arcsec}\\
Instrument & AIA & AIA\\
Passband & 171,193,211  & 171,193,211\\
Exposure normalize & True & True\\
Cadence & 12 sec& 12 sec \\
\enddata
\end{deluxetable*}


\begin{figure*}[ht!]
\centering
\includegraphics[width=\linewidth]{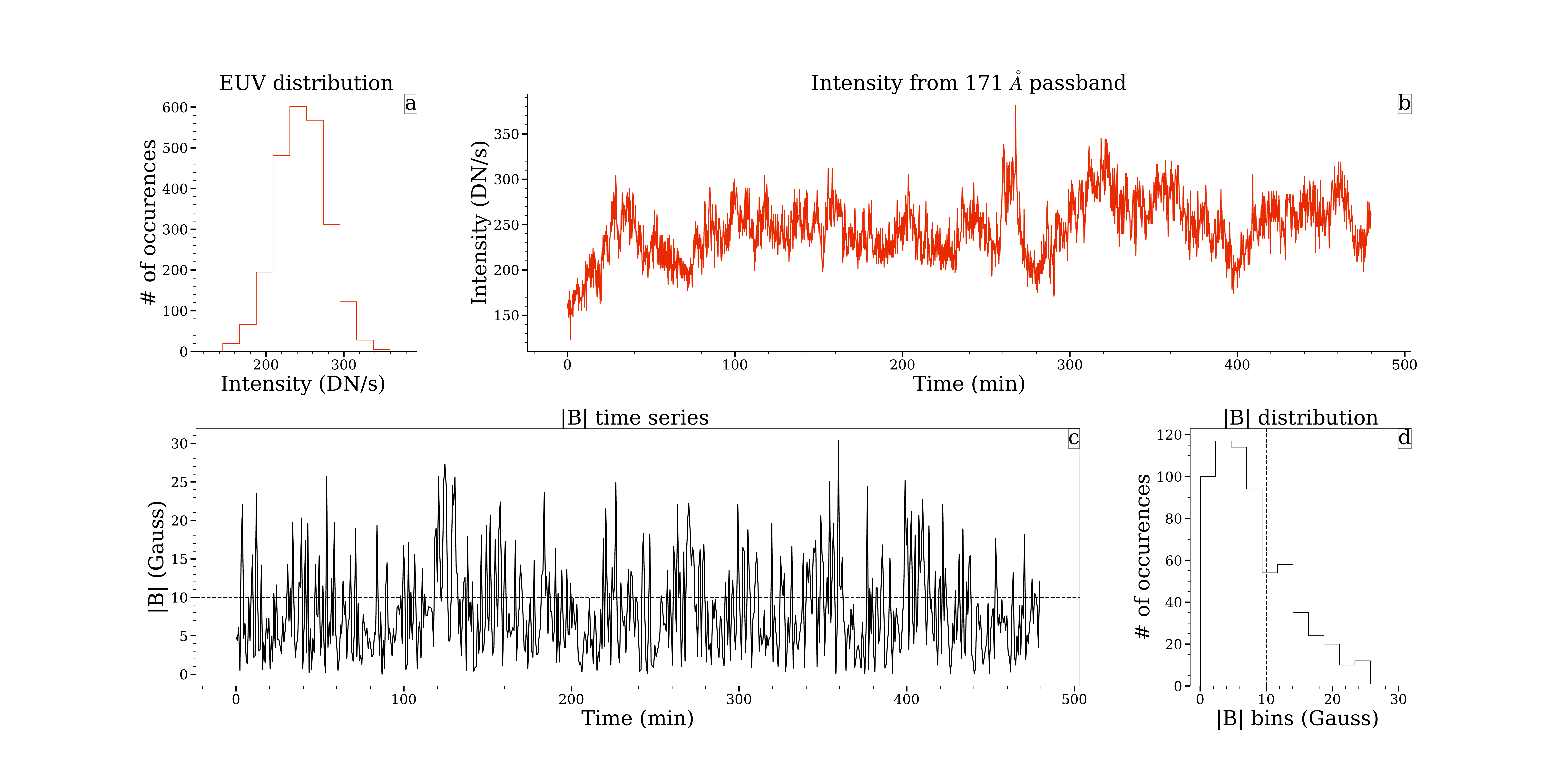}
\caption{Intensity and Magnetic field intensity time series of 1 pixel from the FOV of DS 1, and their corresponding distributions, as labelled. The 10 G noise level has been indicated in (c) and (d). }
\label{fig:1pixdistr1}
\end{figure*}
\begin{figure*}[ht!]
\centering
\includegraphics[width=\linewidth]{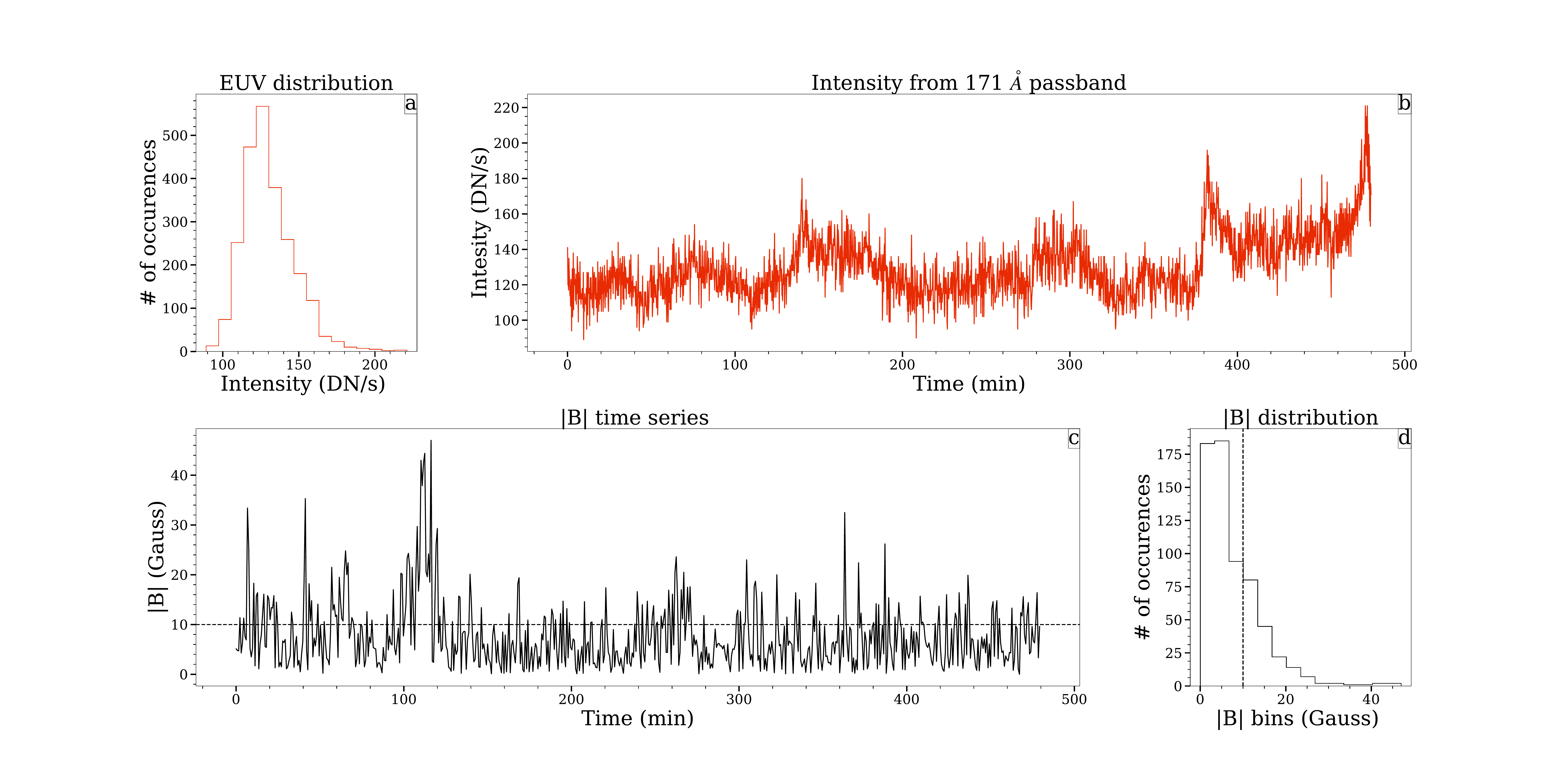}
\caption{Same as Fig.~\ref{fig:1pixdistr1} but for DS2.}
\label{fig:1pixdistr2}
\end{figure*}

\vishal{To study the distribution of the intensity, for each pixel we create light curves of intensity in all three passbands. We plot sample light curves from both the datasets, for one passband, and their corresponding distribution in Figs.~\ref{fig:1pixdistr1} \& \ref{fig:1pixdistr2} for a single pixel. Note that we also show time series of magnetic flux density of the corresponding pixel taken from the Helioseismic and Magnetic Imager \citep[HMI; ][]{HMI} on board SDO corresponding to the AIA Field of View (FOV). The LOS magnetograms are obtained by HMI at approximately 45~s cadence with a pixel size of 0.5{\arcsec}. We map the HMI data to the same plate scale as that of AIA. The error in HMI LOS measurements is estimated to be $\pm$ 10G\thirdround{~\citep{yeo_10gauss_bfielderror,Couvidat2016_HMIErr}}, and this is depicted in the figure as the black horizontal line.} 
The time series for intensity and magnetic flux density are shown in panels b and c. The distributions are shown in panels a and d, respectively, which demonstrate the log-normal distribution, as was previously observed by \citet{PSModel} in SUMER observations and \cite{safari1} for AIA observations. 

\section{Noise Characterization of the light curves} \label{sec:noise}

The observed light curves, shown in Figs.~\ref{fig:1pixdistr1} and ~\ref{fig:1pixdistr2}, have inherent noise, which is essentially dominated by photon shot noise. To mitigate this, while preventing over-smoothing (and thus averaging over real events), we use a procedure that we call \texttt{Finding kneemo}. 

\texttt{Finding kneemo} is based on existing knee analyses performed in Machine learning. Broadly, the goal of the algorithm is to monitor a performance metric against the free parameter, which, in our case, is the size of the smoothing window. The window size for which we observe a drastic improvement in performance metric is taken as the box-car window. The change point is generally known as ``the knee''.

\begin{figure}[h!]
    \centering
    \includegraphics[width=0.5\linewidth]{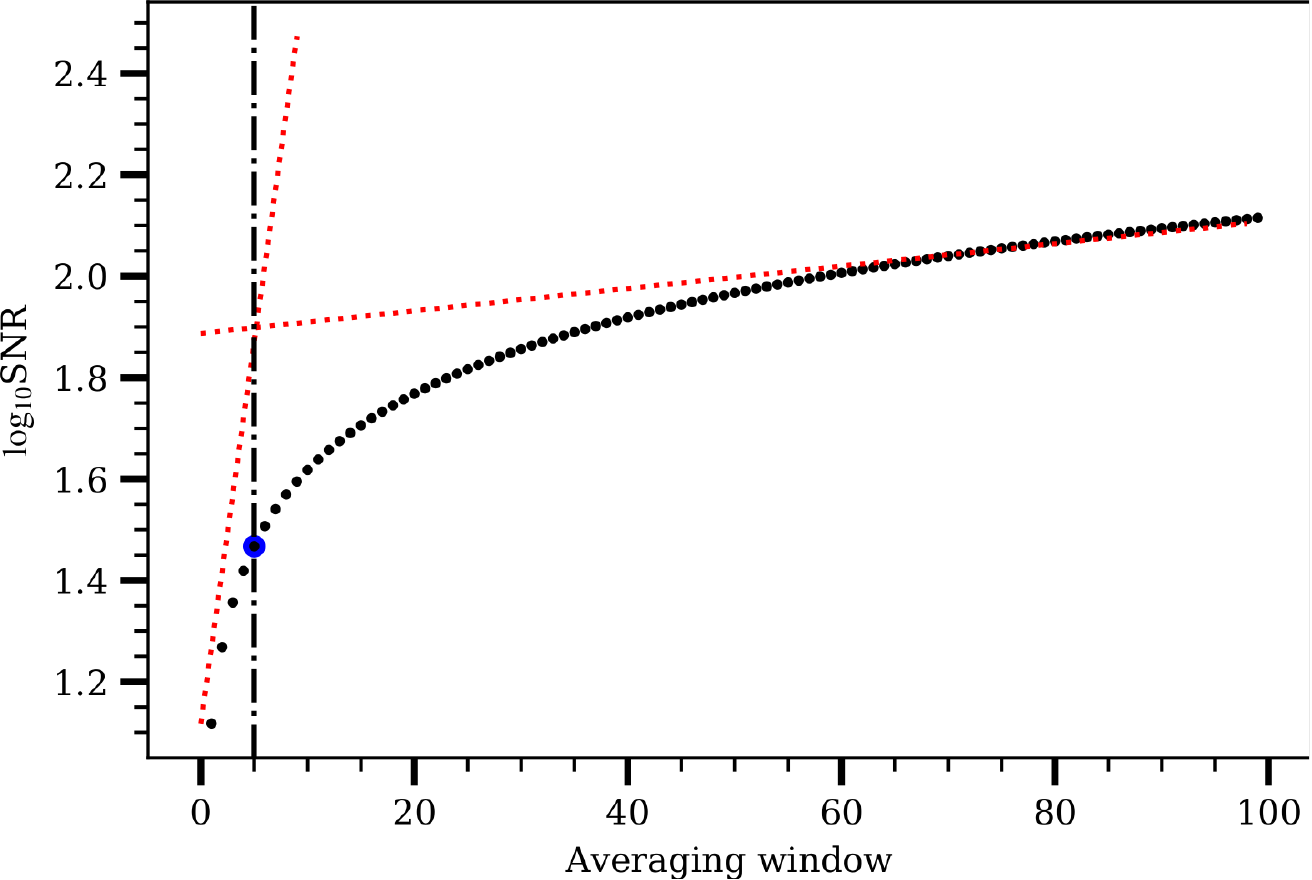}
    \caption{SNR variation with box-car window. The black dots represent the variation, while the red dotted lines represent the asymptotes. The vertical black line shows the approximate point of intersection (marked as blue)}
    \label{fig:snrplot}
\end{figure}

\begin{figure}[ht!]
    \centering
    \includegraphics[width=0.5\linewidth]{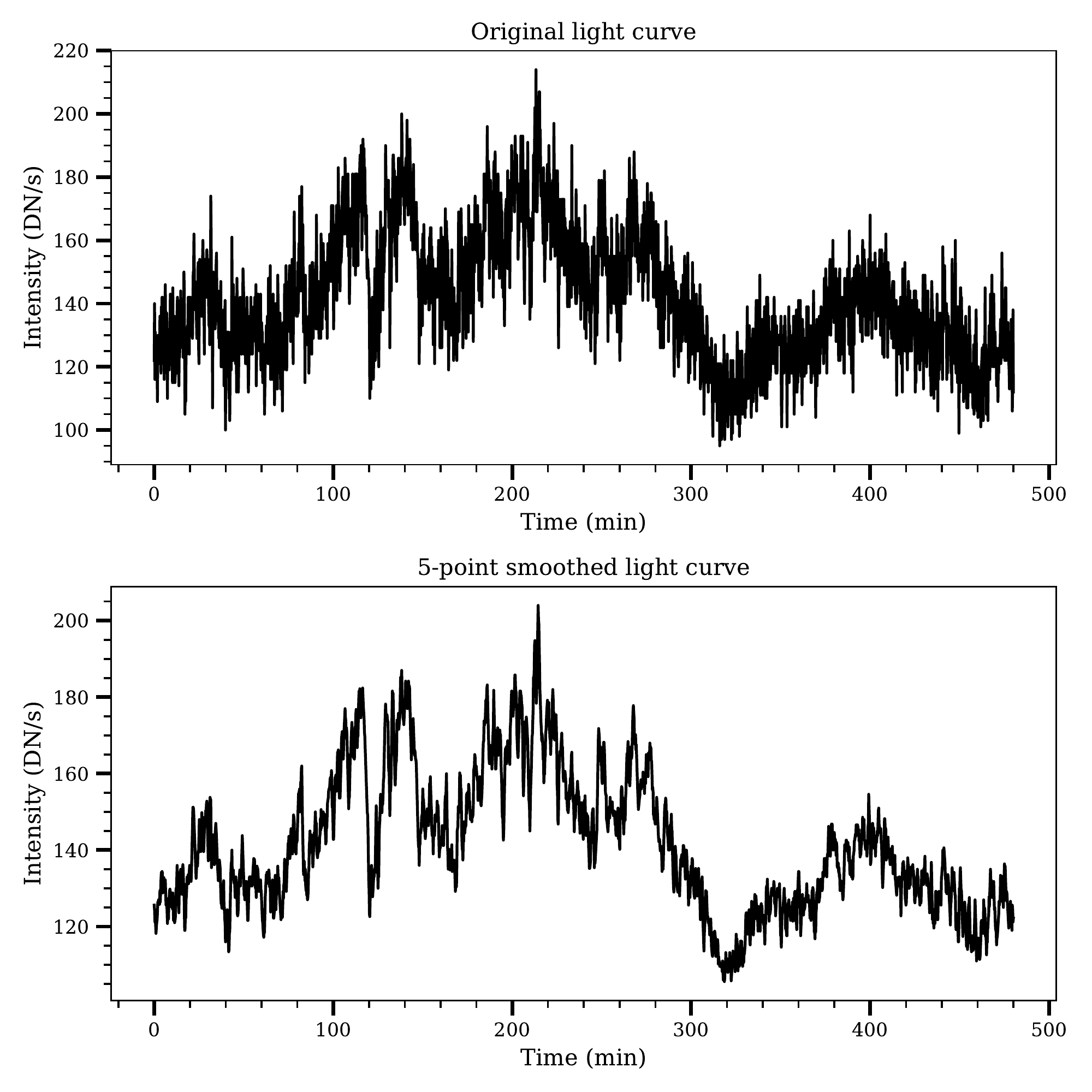}
    \caption{Comparison of original and smoothed light curves by \texttt{Finding kneemo}. }
    \label{fig:snrsmooth}
\end{figure}

The knee determination is extremely qualitative, though some methods exist which quantify this well \citep[see, for example][]{Kneemo}. In our analysis, we consider a random light curve for a pixel in our data set, along with its error time series, which is obtained from \texttt{aia\_bp\_estimate\_error.pro}. We smooth light curve using box-car of box size varying between 1 and 100, and obtain its modified SNR (Signal-to-Noise Ratio). We then plot the obtained SNR against the box size in Fig.~\ref{fig:snrplot}, along with the asymptotes of the SNR, and find their point of intersection. This point is then selected as the box-car window size. From Fig.~\ref{fig:snrplot}, we find that the asymptotes intersect at box-car size of 5 time-points. Therefore, we use this value to enhance SNR. Note that we have run this analysis on several lightcurve within our dataset and have found a consistent result. Thus, we have performed box-car averaging with box size of 5 time-points for all the light curves in our dataset. An example plot with the original and the smoothed light curve is shown in Fig.~\ref{fig:snrsmooth}.

\section{Methods} \label{sec:methods}
\subsection{Nanoflare model} \label{subsec:nanoflare_model}
\begin{figure*}[htpb!]
    \centering
    \plotone{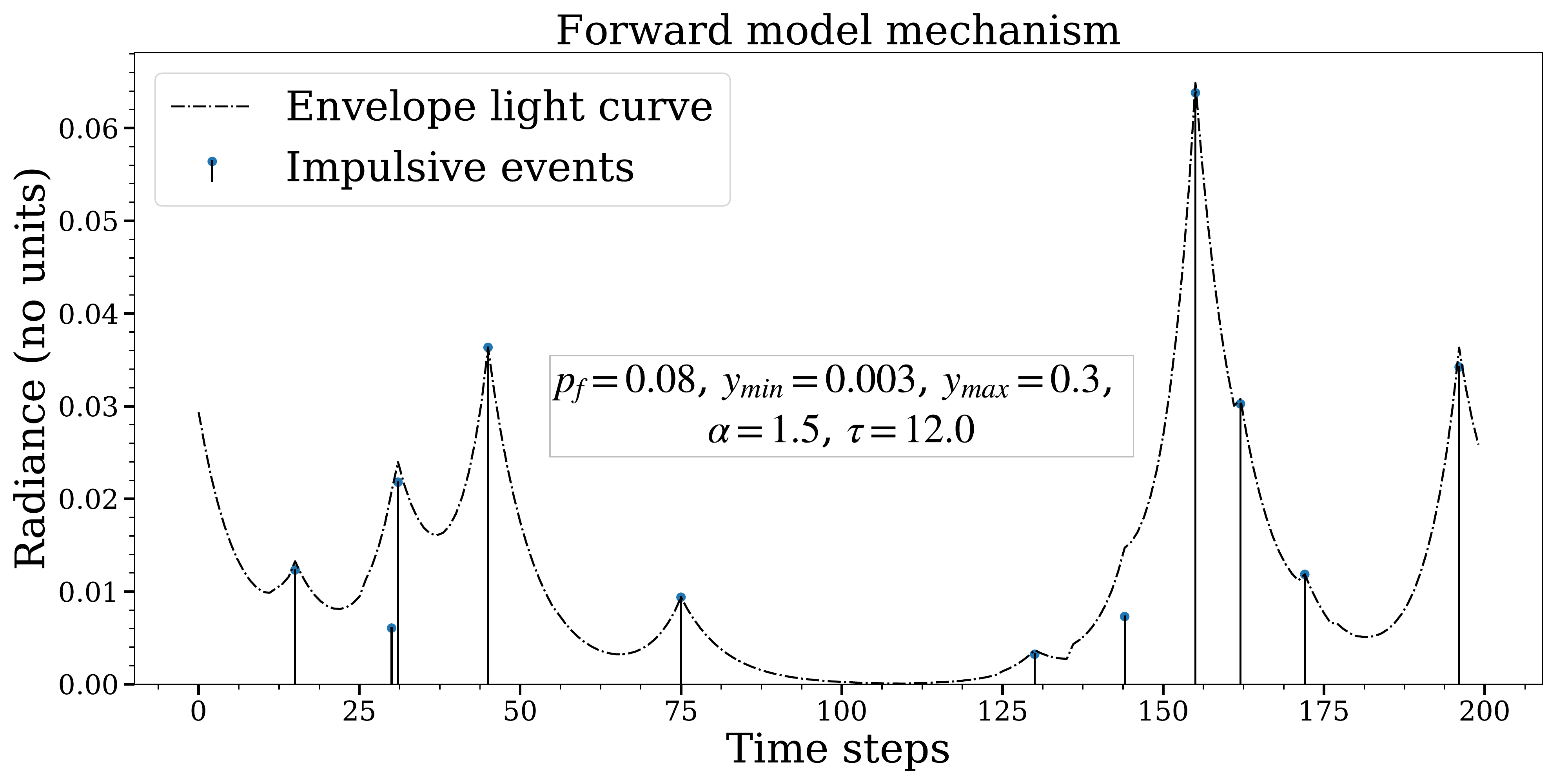}
    \caption{\vishal \thirdround {An example showing} light curve generation using the Nanoflare generation model, similar to Fig. 2 of \cite{PSModel}.}
    \label{fig:nanoflaregen}
\end{figure*}

The forward model employed here, as mentioned earlier, is the \texttt{PSM}. This is an empirical model based on two key observations, i.e. log-normal distribution of spatial and temporal distribution of QS intensities~\citep{ps_2001}, and power law distribution of energies from flares to micro-flares \citep[see, e.g.][]{aschbook}. The algorithm may be summarised by asking the following questions: 

\begin{enumerate}
\item What is the probability of a flare to occur at a given time step?
\item If a flare is meant to occur at the given time step, what would its peak energy be?
\item How long will the flare last once it has occurred, i.e. the duration of the flare?
\end{enumerate} 

In this model there are 5 free parameters: the event or flaring frequency ($p_f$), i.e., the probability of a flare to occur at a given time; the duration of the individual flare event ($\tau$); the power law slope ($\alpha$) and the minimum ($y_{min}$) and the maximum ($y_{max}$) energy that is allowed for an individual flare events, which provides the bounds of the power law. \vishal{An example} simulated light curve\vishal{, depicting the formation of light curve from individual events} is shown in Fig.~\ref{fig:nanoflaregen}. \vishal{It may be seen from here that given only the envelope light curve, the individual events may not be inferred.} We note that the simulations are performed over a large parameter space (see Table~\ref{tab:param_rng}), while fixing $y_{min}$ and $y_{max}$, to perform the inversion of light curves from all the three passbands with the same inversion model.

We generate the light curves for length of $5L+1600$, where $L$ is the length of the light curve (2400 in our case), and reject 800 samples from either side to remove boundary effects. The remaining light curve is folded $5$ times to get a final light curve of length $L$\vishal{\footnote{Folding essentially divides up the light curve in 5 equal chunks of length L and gets the average curve from these chunks}}. This is done to \vishal{minimize} the effects of the initial seed for random number generators. The observed light curves are normalized by their median values, following \citet{safari2}. \thirdround{Hence, the radiances reported from hereon have no associated units.}

\begin{deluxetable*}{ccc}[!ht]
\tablecaption{Simulation grid parameters. Note that all parameters are in code units. \label{tab:param_rng}}
\tablewidth{0pt}
\tablehead{
\colhead{Parameter} & \colhead{Range} & \colhead{Stepsize}
}
\startdata
$p_f$  & $[0.05,0.95)$ & steps of 0.05 \\
$\alpha$ & $[1.1,3.0)$ &steps of 0.1 \\
$\tau$ & $[1,100)$ & steps of 2.0 \\
$y_{max}$ & 0.3 & \\
$y_{min}$ & 0.03 & \\
\enddata
\end{deluxetable*}
\subsection{Inversion}\label{subsec:inversion}

We perform the inversion using 1-D CNN. In this approach, generally, there are convolution layers followed by an activation layer. The activation, as we have seen in Sec.~\ref{sec:intro}, is a non-linear function which forms the core of complex learning ability of any NN. We use the function \texttt{Elu} as defined in Eqn~\ref{eqn:elu} as activation for all layers except the last, since \texttt{Elu} enables the network to train faster and generalize better~\citep{elu}. For the final layer, we have no activation since this is a regression problem mapping to a continuous variable.

\begin{eqnarray}
Elu := &  \left\{
\begin{array}{cc}
e^x-1 & \text{ , if $x\leq0$}\\
x & \text{ , else}
\end{array}\right.
\label{eqn:elu}
\end{eqnarray}

\begin{figure*}[!ht]
\plotone{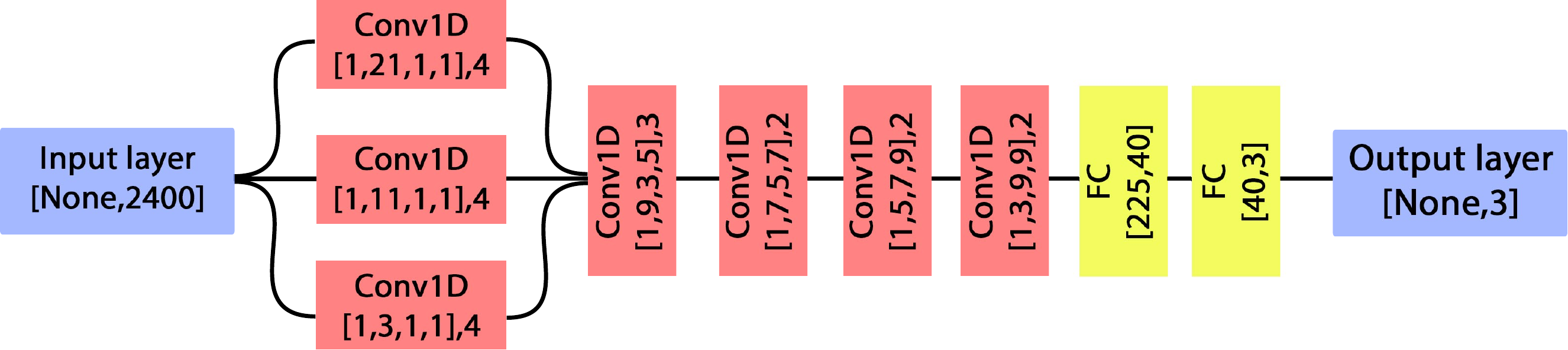}
\caption{CNN architecture used in this work. The blue boxes indicate input/output layers, and other colours indicate trainable layers. The tensor shapes are given in square brackets as [ ], and the number outside, for the Convolution layers is the stride size. FC denotes the fully connected layers.} \label{fig:nnModel}
\end{figure*}

A graphical representation of the model architecture is shown in Fig.~\ref{fig:nnModel}. As depicted in the figure, there are input/output layers, convolution layers (where we implicitly assume the activation function to be present), and fully connected (FC) layers. In the tensor shape, `None' is generally used to denote a variable size, which in this case represents number of light curves to be inverted during a single forward pass. The convolution layers (marked in red) are given a 4-dimensional shape, representing [height, width, input channel, output channel], and the stride size given by an integer. Note that ``channel" here is not to be confused with AIA passbands, and that since we have used a 1-D signal, the height is set to 1. After a suitable number of convolutions, the array is unrolled to 1D, and captured by fully-connected layers (marked in yellow). For more information on general CNN and training see \citet[][Chapter 9]{Goodfellow-et-al-2016}.

\subsection{Data preparation and training}\label{subsec:dataprep}

We have divided the simulated light curves into training set (80\%) and testing set (20\%). However, before feeding in any data to the CNN, it must be prepared appropriately to make sure all the parameters are of the same scale. The training set parameters are rescaled between 0 and 1, and the testing set parameters are rescaled using the training statistics, as is the standard procedure in machine learning. All light curves are also rescaled between 0 and 1.

We train the CNN by feeding in the training set light curves, and obtain three free parameter set ($p_f$, $\tau$ and $\alpha$). This obtained parameter set is them compared with the original target parameter set using an error metric, which is used to update the kernel values of the CNN. The error metric is the sum of the two terms defined as: 
\begin{eqnarray} \label{metric}
\mathcal{L}_1(\hat{\bold{x}},\mathbf{x}) := \Sigma_i \left|x_i-\hat{x}_i\right|, 
\mathcal{L}_2(\hat{\mathbf{x}},\mathbf{x}) := \Sigma_i \left(x_i-\hat{x}_i\right)^2 ,
\label{eqn:error}
\end{eqnarray}

where $\mathbf{x}$ is the target parameter set, $\hat{\mathbf{x}}$ is the predicted parameter set, and the summation is performed over all observations, and all parameters. 

There are certain ``free parameters" that needs to be fixed while developing any NN. These free parameters are called hyper-parameters, while the trainable parameters of the NN are called the ``weights" of the NN. For training the CNN, we have used the ``Adam Optimizer"~\citep{kingma2014adam}, which is a stochastic optimization algorithm. The ``size" of update at each step is controlled by the hyper-parameter called learning rate. 

Overfitting is a serious issue in NN training whereby the model starts fitting the noise in the model, and stops generalizing. This may result in erroneous results and interpretation of inversion. To prevent over fitting, we use dropout \citep{hinton2012improving}, which essentially switches off neurons randomly with a fixed probability for every forward pass. The training hyper parameters are summarized in Table.~\ref{tab:train_param}.

\begin{deluxetable*}{cc}
\tablecaption{Training Hyperparameters \label{tab:train_param}}
\tablewidth{0pt}
\tablehead{\colhead{Hyperparameter} & \colhead{Value}}
\startdata
Cost function & $\mathcal{L}_1$(prediction,target)+$\mathcal{L}_2$(prediction,target) \\
Optimizer & Adam Optimizer with default values\\ 
Learning rate & $1e-3$\\
Dropout rate & $0.2$\\ 
No. of iterations & $3000$\\ 
\enddata
\end{deluxetable*}
\subsection{Uncertainty estimation}\label{subsec:uncert}

A CNN generates a single prediction for a given forward pass. However, in general, there exist two kinds of uncertainties -- namely Epistemic and Aleatoric -- associated with any such predictions ~\citep{kendall2017uncertainties}. Epistemic uncertainty relates to model uncertainty due to unexplored weight space of the neural network. Aleatoric uncertainty relates to the inherent uncertainty in the target parameter values. In our study, there is no Aleatoric uncertainty because the parameters are defined by us. Hence, we have only Epistemic uncertainty.

\vishal{Deficiencies in model training, resulting in unexplored weight space, can occur if not enough data is provided during model training. Hence, this effect can be simply minimized by increasing the size of training set. However, the epistemic uncertainty measure, while informing us about the deficiencies in fitting, may also inform us about any outliers in the dataset. Throughout the analysis in this work, the \texttt{PSM} is assumed to be the ground truth, i.e, it fully describes the observed light curves to infer parameters. This is obviously never the case with any model. Hence, departures of the observations from simulations, where the \texttt{PSM} does not fully explain the given observation, would behave as outliers. Hence higher uncertainties associated with the parameters inferred from the observations tells us either there are deficiencies in model fitting in certain regimes, or the light curve is not explained well by the \texttt{PSM}. However, we note that it is practically impossible to disentangle these effects \citep[see e.g.,][]{kendall2017uncertainties}. }

The epistemic uncertainty may be estimates by application of dropout~\citep{hinton2012improving,diazbaso}. In addition to being used to prevent overfitting, Dropout can also be used to create perturbations, and obtain the variability in the predictions~\citep{gal2016uncertainty}. Since the neurons switched off in every forward pass are random, we perform a Monte Carlo forward pass to obtain multiple realizations of the CNN, and present the mean, standard deviation from the passes. Thus, the error bars reported for the estimated parameters of an individual light curve are the standard deviation obtained from Dropout. 
\subsection{Inversion performance}\label{subsec:nn_perf}
\begin{figure*}[htbp!]
\includegraphics[width=0.3\linewidth]{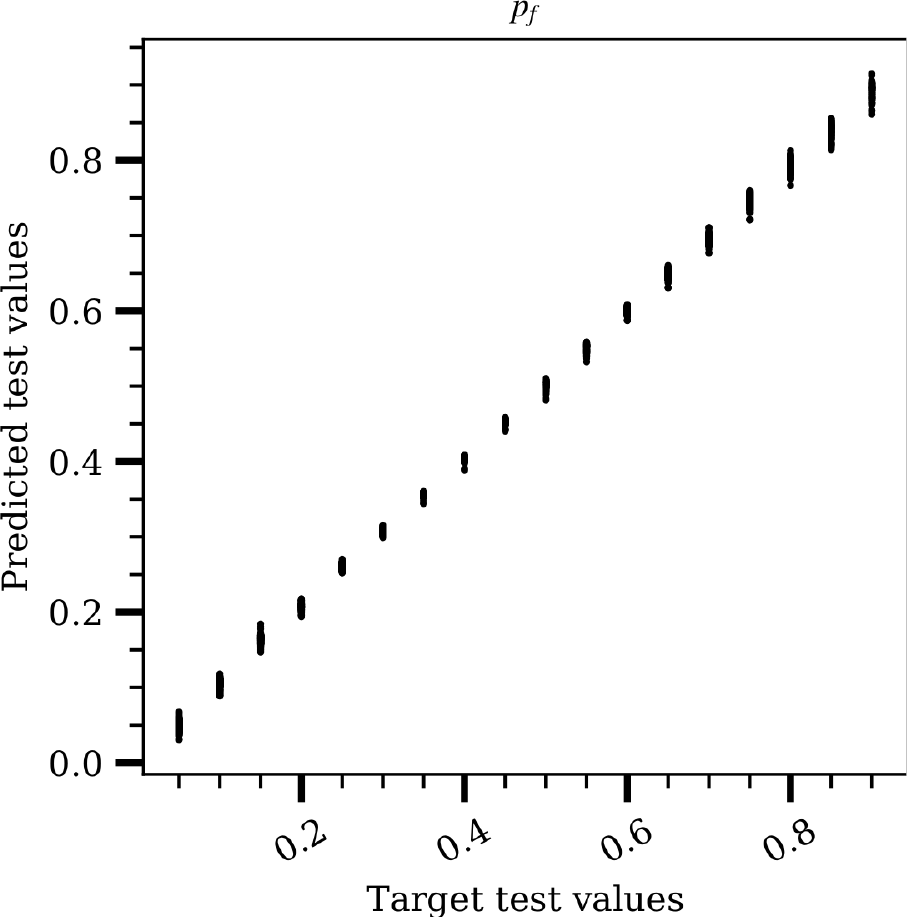}
\includegraphics[width=0.3\linewidth]{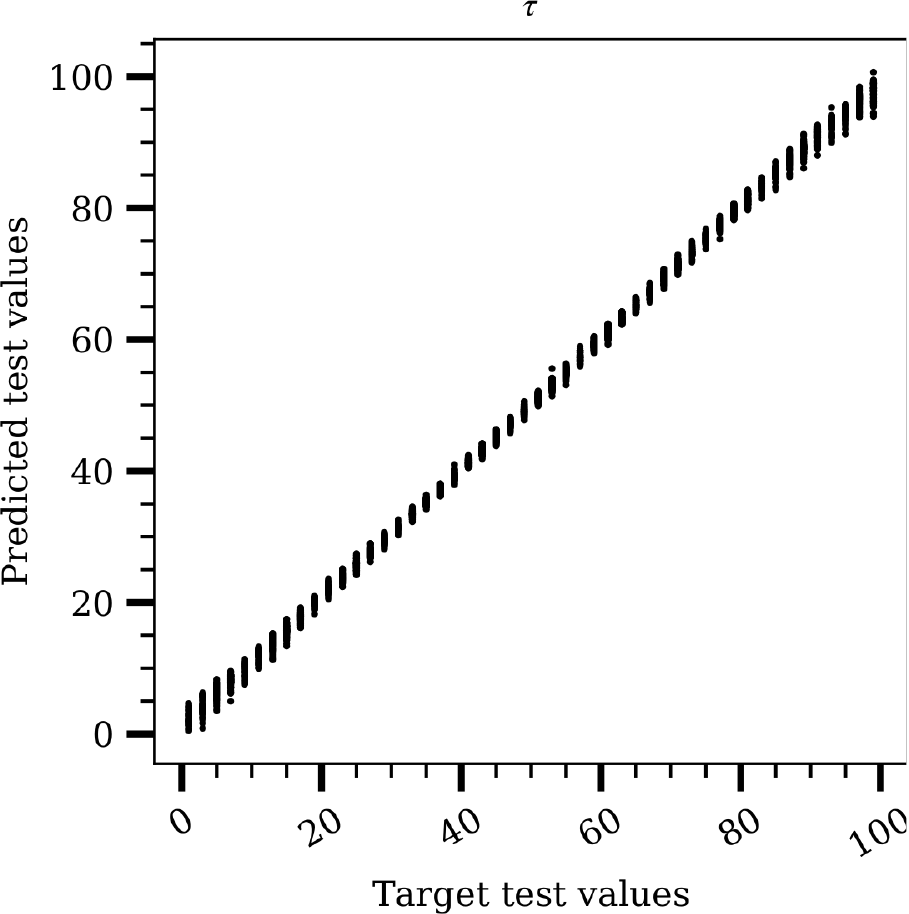}
\includegraphics[width=0.3\linewidth]{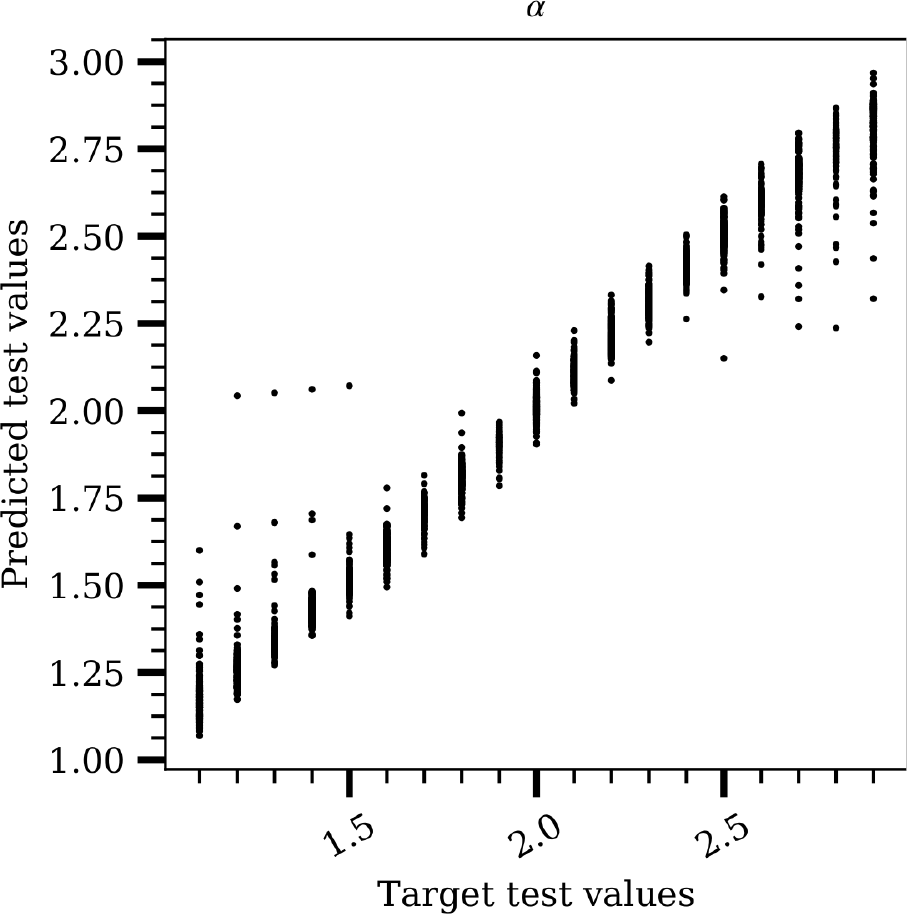}
\caption{Correlation plot of target and predicted parameter values from our CNN, for $p_f$(left), $\tau$ (center) and $\alpha$ (right). This quantifies the generalizability of the CNN from the training set.}
\label{fig:nn_params}
\end{figure*}

To asses the performance of our CNN, in Fig.\ref{fig:nn_params}, we display scatter plots between the target and predicted values of $p_f$ (left panel), $\tau$ (middle panel) and $\alpha$ (right panel). As can be readily noted, the predicted values lie very close to the target values, thereby validating the performance of our network on a test set. This may be quantified using the coefficient of determination ($R^2$) defined as:

\begin{equation}
R^2 := 1-\frac{\Sigma_i \left(x_i-\hat{x}_i\right)^2 }{\Sigma_i \left(x_i-<x>_i\right)^2 },
\end{equation}

where $<x>$ represents the mean of target set, $\bold{x} = \{x_i\}$ represents the target values, and $\hat{\bold{x}} = \{\hat{x}_i\}$ represents the predicted values. In this case, $i$ corresponds to number of points in the test set, i.e. $R^2$ is computed separately for each target parameter. The $R^2$ values are 0.990, 0.999 and 0.97 for $p_f$, $\tau$ and $\alpha$, respectively, showing excellent performance of our network. Our network is now ready to be fed in with the observed intensity light curve.

\section{Results} \label{sec:results}

Now we discuss the application of the network on the observed light curves. We first discuss the results obtained for a single light curve in \S\ref{fig:singleLC}. Then in \S\ref{subsec:multi}, we move on to discuss the results obtained for all light curves obtained for all the three AIA passbands. We next explore the various correlations between our parameters and perform an analysis of the involved energetics in \S\ref{subsec:energy}.

\subsection{Application of the CNN on a Single Light curve} \label{fig:singleLC}

For representative purpose, we choose the intensity light curve for a random pixel from both DS1 and DS2 and obtain the corresponding simulated light curve. \newref{It is important to note that since \psm~is a statistical model which generates a representation of the observations ``statistically'', one should not perform a point by point comparison of the simulations with the observations. A simple change in the seed of the random number generator can change the exact times when events occur. Furthermore, since the amplitude of events is sampled from a distribution, the random seed value can also change the event amplitude at particular times. However, these seeds cannot change the overall statistical properties of the light curve. Thus, a comparison of the observation and simulation must be done using statistical properties of light curves (e.g intensity distribution, frequencies showing enhances power), rather than a pointwise comparison of light curves. Once a good representative simulated light curve is obtained, the corresponding parameter set is taken to characterise the observed light curve.}

\begin{figure*}[htbp!]
\plotone{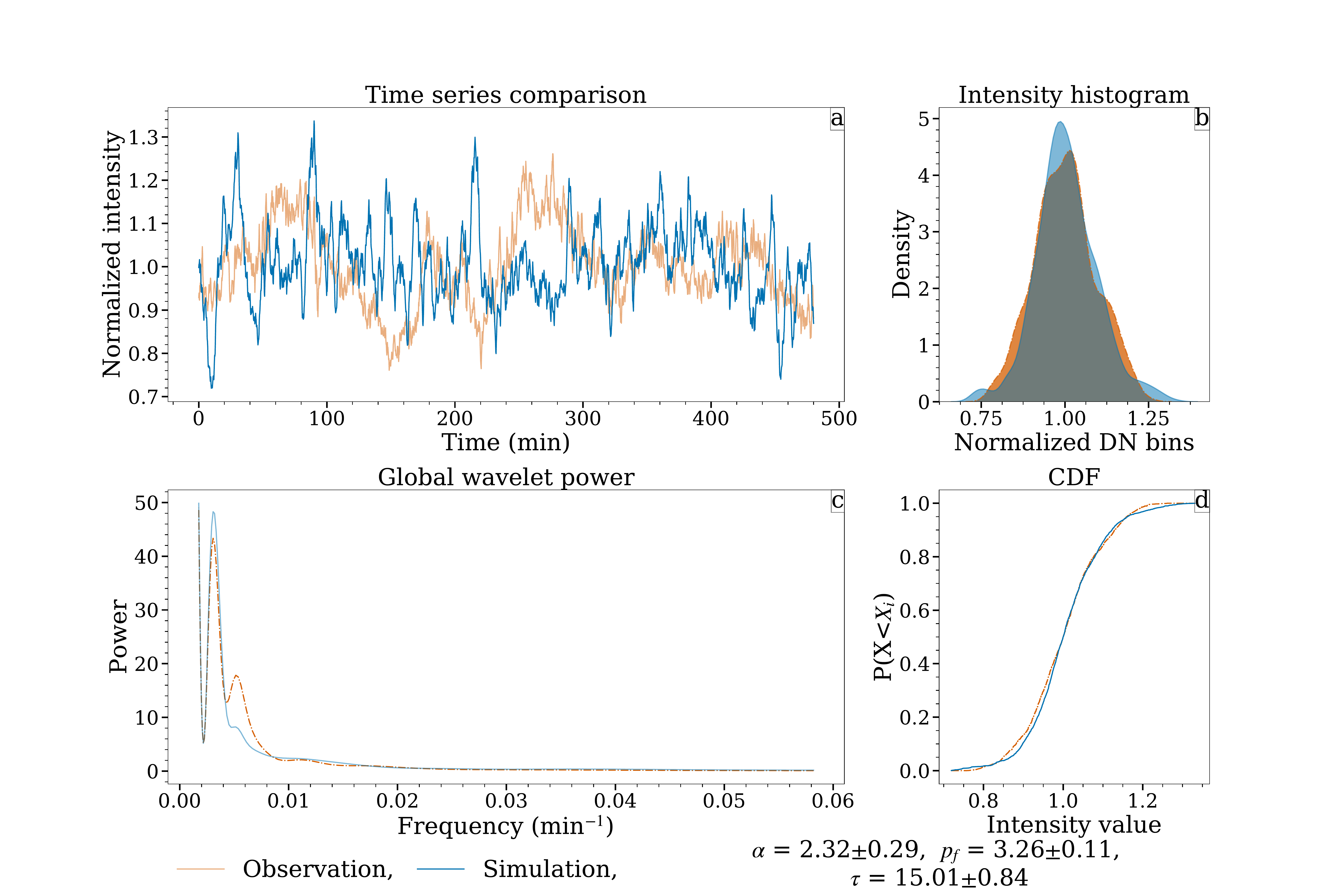}
\caption{The comparison of a representative light curve obtained for 171~{\AA} passband from DS1 with a simulated light curve. Observations are shown in orange translucent and simulations are shown in blue. Panel a: Normalised observed and simulated light curves; Panel b: KDE of observed and simulated light curves; Panel c: Global Morl\'et power for observation and simulations; panel d: Cumulative Distribution Function (CDF) comparison of simulation and observation.} \label{fig:replc_ds1}
\end{figure*}

\begin{figure*}[htbp!]
\plotone{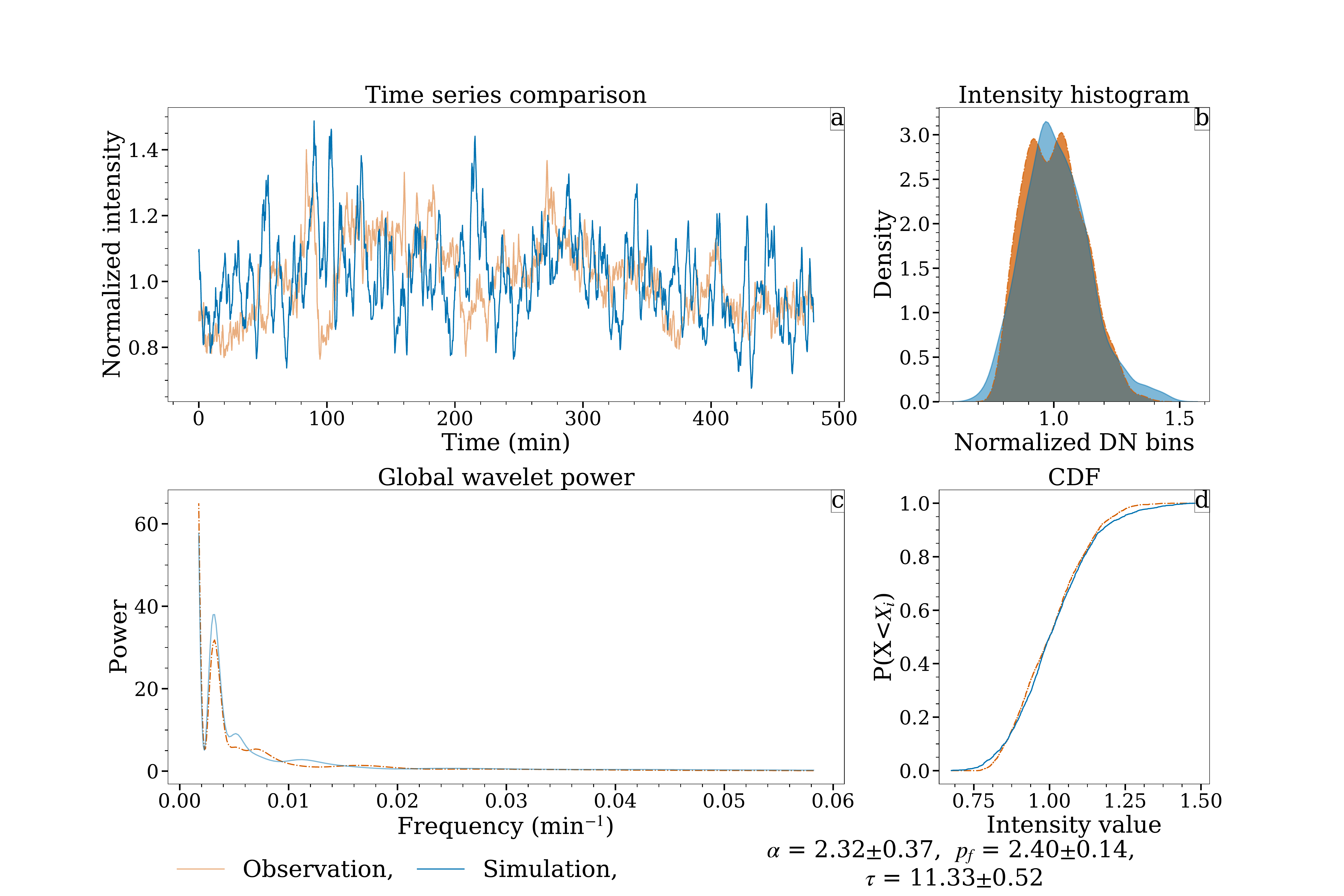}
\caption{Same as Fig.~\ref{fig:replc_ds1}, but for DS2.}
\label{fig:replc_ds2}
\end{figure*}

In Figs.~\ref{fig:replc_ds1} and \ref{fig:replc_ds2}, we show the comparison between observed (orange) and simulated (blue) light curves (panels a), \newref{Kernel Density Estimation (KDE)} of intensity distribution (panels b), the Global Morl\'et power spectra (panels c) and the cumulative distribution function (CDF; panels d). Note that both the observed and simulated light curves are normalized by their median. The parameter sets denoted for the simulated light curves are the mean values of the obtained parameter distribution by performing 1000 Monte Carlo simulations, and are denoted at the bottom of the figures.\newref{The KDE can be understood to be essentially a continuous extension of histogram~\citep[see eg.][]{chen2017_kde}}. \vishal{Note that $p_f$ and $\tau$ are defined as per time step in Fig.~\ref{fig:nn_params}, and may be converted into real units as $$p_f(\text{per min}) = \frac{p_f(\text{inferred})}{\text{Cadence(min)}},$$ and
$$\tau(\text{min}) = \tau(\text{inferred})\times\text{Cadence(min)}.$$ The $p_f$ denoted henceforth is given as number of events per minute, while $\tau$ is given as timescale in minutes. The $\alpha$ remains a dimensionless parameter.}


From panels a of Figs.~\ref{fig:replc_ds1} and \ref{fig:replc_ds2}, we can see an excellent statistical correspondence between the observed and simulated light curve. This is corroborated by the match in their corresponding KDEs (panel b) and CDFs (panel d). Furthermore, the Morl\'et wavelet power spectrum shows peaks at corresponding frequencies for both the observed and simulated light curve. These results confirm that the Inversion model was able to learn both the time series and distribution properties corresponding to the 3 free parameters. We further emphasize that value of $\alpha$ inferred in these two cases is $\ge2$, which in turn suggests that events with smaller energy are dominantly responsible for generation of radiance of these particular examples. 

From Figs.~\ref{fig:replc_ds1} and ~\ref{fig:replc_ds2}, we find a clear relation between the goodness of representation of simulated light curve (using CDF and Morl\'et Power) and the spread of parameters obtained by Monte Carlo simulations. Consider the percentage uncertainty (i.e, uncertainty/mean value) -- we find this quantity is approximately $3\%$ in $p_f$ for DS1, $6\%$ in $p_f$ for DS2; $\sim5\%$ in $\tau$ for DS1 and $\sim4.5\%$ DS2; and $\sim12\%$ for DS1 and $\sim15\%$ for DS2. Thus, the Inversion model is more certain of the parameters of DS1 than DS2, which is also reflected in the relative mismatch of Morl\'et power between the observation and simulation for DS2 over DS1, at the first two peaks (note the difference in y-axis limits in panels c). Thus, such an uncertainty measure, along with the Monte Carlo forward pass, can help us explain which parameters are strongly influencing the quality of a given inversion assuming \texttt{PSM} as the ground truth.

\subsection{Multi-light curve - multi-passband analysis} \label{subsec:multi}

Since our network gives reliable results for the light curve obtained for a random pixel in both the data set, we now take all of our light curves ($331967$ light curves per passband), for the three AIA passbands, and pass them through the network to obtain the relevant parameter set for each light curve. Due to operational constraints, we perform only 100 Monte Carlo forward passes in this case. We emphasize that the obtained parameter set for 100 and 1000 Monte Carlo forward passes, are statistically same for a limited, handpicked set of representative light curves. 

\begin{figure*}
\centering
\includegraphics[width=0.3\linewidth]{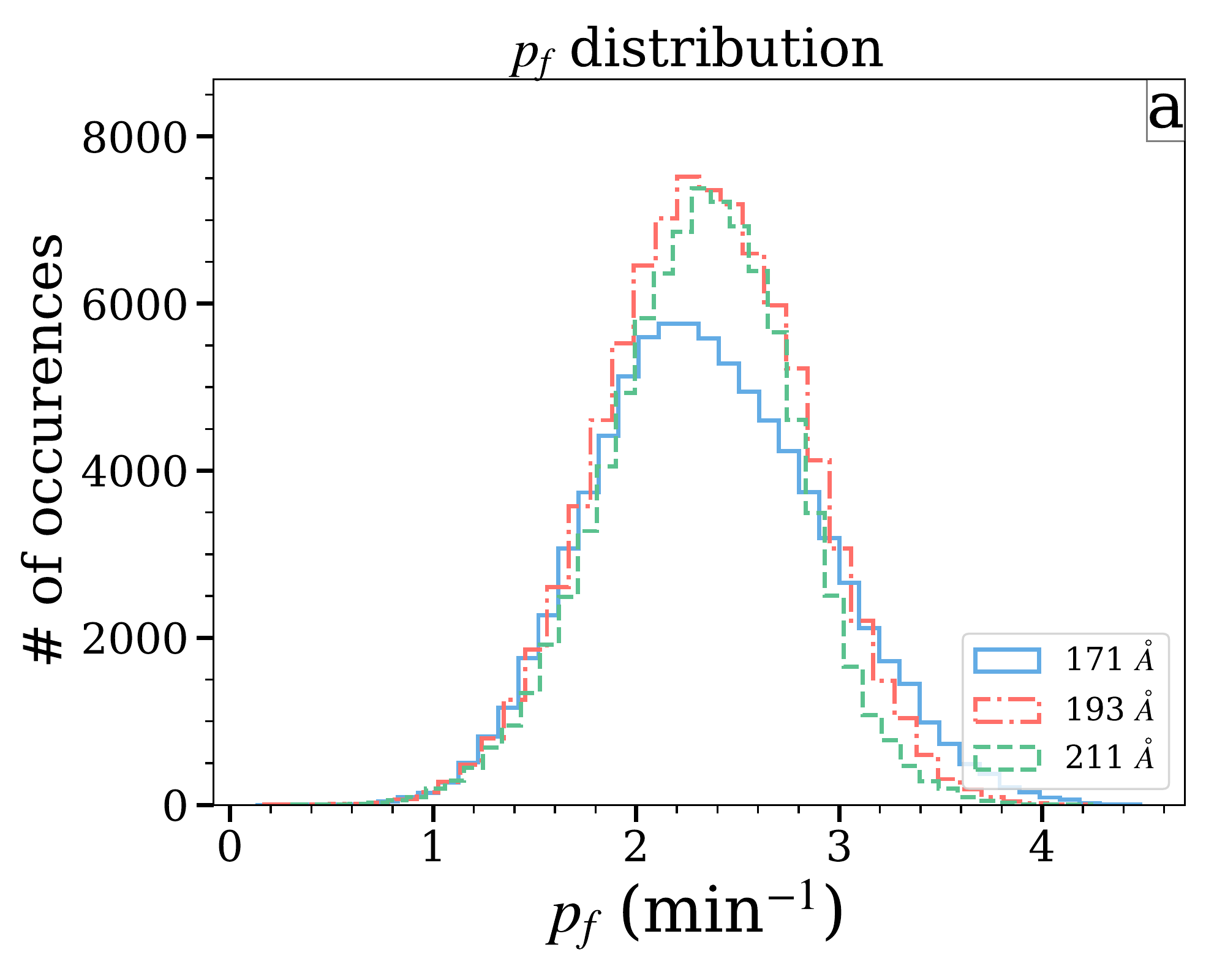}
\includegraphics[width=0.3\linewidth]{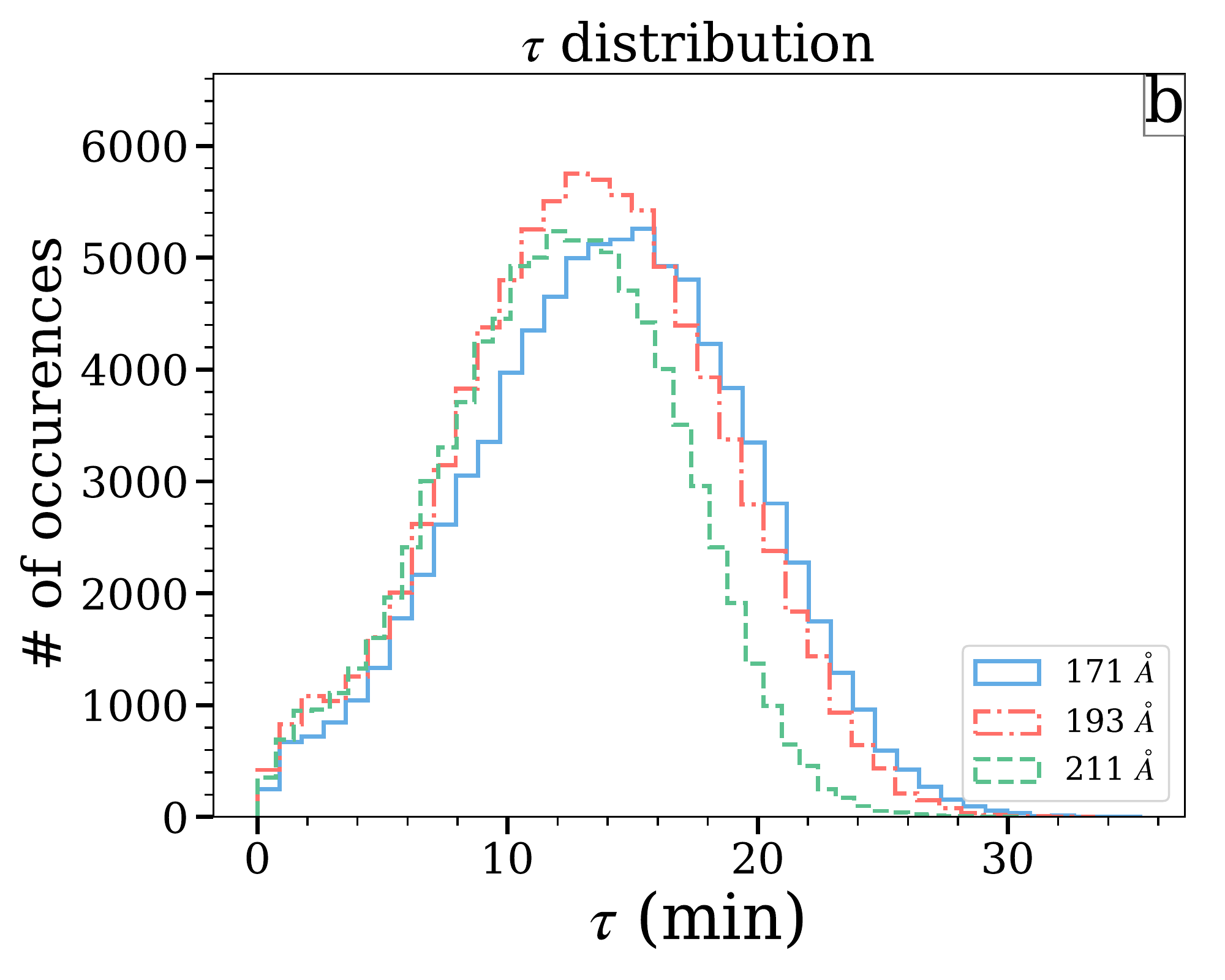}
\includegraphics[width=0.3\linewidth]{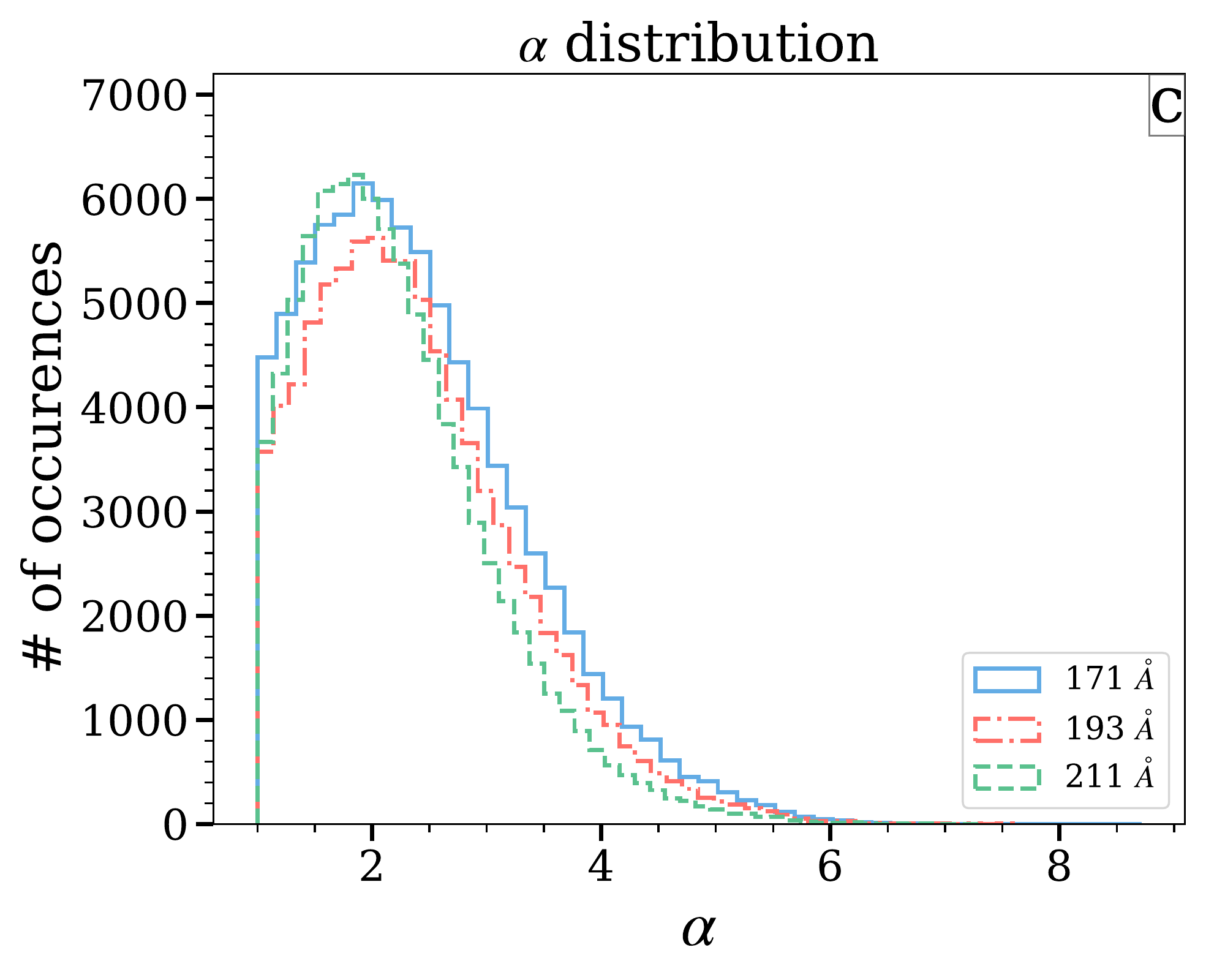}
\caption{Distribution of inferred parameter set for $p_f$ (panel a), $\tau$ (panel b) and $\alpha$ (panel c) over both the datasets. The colours are distributed as blue (171~{\AA}), red (193~{\AA}) and purple (211~{\AA}).} \label{fig:FOVdistr}
\end{figure*}

For both of our data set, we first divide each light curve by its median value, rescale between 0 and 1, perform the Monte Carlo forward pass through the CNN, and obtain the mean parameter set. Finally, we concatenate the parameter set across the whole field of view for both data set separately for each AIA passband to improve our statistics. This concatenation can be done since all light curves are from QS regions and are evolving independently. Fig.\ref{fig:FOVdistr} displays the distribution of flaring frequency $p_f$ (panel a), duration $\tau$ (panel b) and power-law slope $\alpha$ (panel c) for this concatenated dataset. The solid blue curves are for 171~{\AA}, dashed-dotted red for 193~{\AA} and dashed green are for 211~{\AA} observations.

The plots reveal that the distribution of all three parameters for all passbands are remarkably similar. The p$_f$ distribution peaks at \vishal{$\sim2.2$ events per minute for 171~{\AA} and at $\sim$2.4 events per minute for 193 and 211~{\AA}, with a range of values between 1 and 4 events per minute. The distribution of $\tau$ peaks near 12 minutes for 211~{\AA}, 14 minutes for 193~{\AA} and 16 minutes for 171~{\AA}, implying a slight temperature dependence. However, we emphasize that since the AIA passbands are multi-thermal, this inference should be taken with caution.}

\begin{figure}[h!]
    \centering
    \plotone{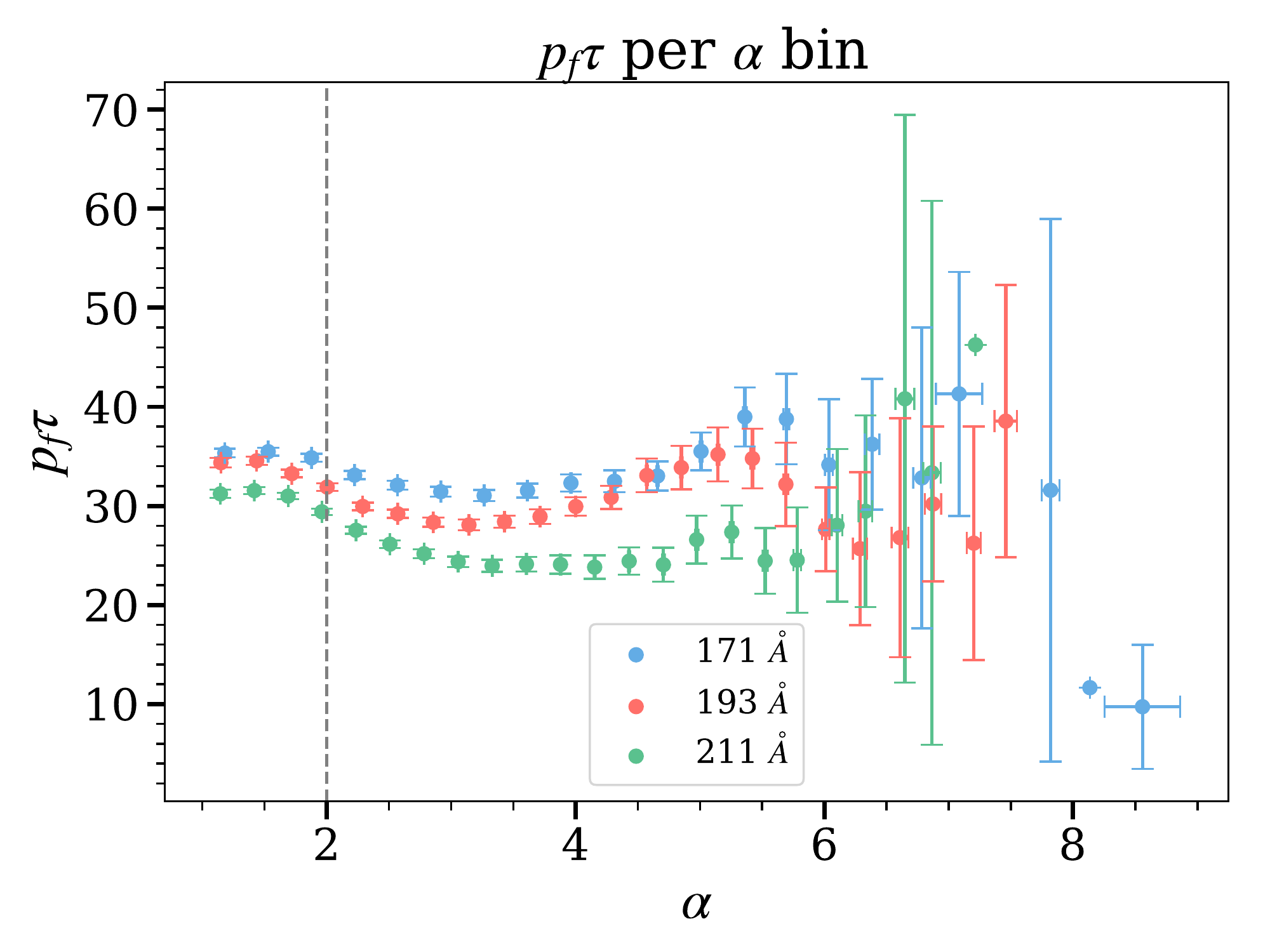}
    \caption{Variation of $p_f\tau$ with $\alpha$. The vertical dashed line marks $\alpha=2$. The errorbars are $3\sigma$ standard erorrs.}
    \label{fig:pftau_alpha}
\end{figure}

The distribution of $\alpha$, plotted in panel c, ranges between $1.0$ to $8$, with peak at $\sim~2.3$ for all three passbands. When we consider the whole set of light curves, we find that $\sim62\%$ of light curves in $171$~{\AA} have $\alpha\ge2$, while the fraction is $\sim61\%$ and $\sim54\%$ respectively for $193$~{\AA} and 211~{\AA} passbands respectively. Thus, we find a reduction in dominance of lesser energy events as we progressively probe the QS in hotter passbands.  We also note that the fall-off for all the three parameters goes from the hotter passband (i.e. 211~{\AA}) falling off first, followed by progressively cooler channels (193 and 171 ~{\AA} respectively).

To further understand any peculiarities exhibited by the QS for the two domains of $\alpha$, we plot the variation of $p_f\tau$ with $\alpha$ in Fig.~\ref{fig:pftau_alpha}. To boost the SNR, we have averaged $p_f\tau$ within a constant bin of $\alpha$. Note that the bin size for averaging is obtained from \cite{doanerule}. The error bars shown are 3$\sigma$ standard error\footnote{Standard error is defined as $\sigma/\sqrt{N}$, where $\sigma$ is the standard deviation in the sample present in the bin, and $N$ is the number of points in the sample}. Following \cite{PSModel}, the factor $p_f\tau$ may be interpreted as the ratio of excitation \vishal{(i.e, intensity generation by $p_f$)} to damping \vishal{(i.e, intensity dissipation by $\tau$) for a given pixel. Here, we investigate the dominance of one over the other.}

The plot reveals that there is almost no change in the excitation to damping ratio for $\alpha<1.8$, and is independent of $\alpha$ in this regime. However, for $\alpha\ge1.8$ the dynamics changes, and we find a reduction in the ratio with increasing $\alpha$. The larger errorbars at the end are due to poor statistics. But even considering the variation till $\alpha=4$, we find a considerable reduction in the ratio with $\alpha$,  presenting an increasing dominance of damping over excitation. Thus, there is either a reduction in excitation, or an increase in damping, or both, which comes into play once the smaller events start dominating radiance generation.

\subsection{Energetics} \label{subsec:energy}

With the parameter set obtained, we can now investigate the involved energetics. A simple way to understand the energetics is from comparing the behaviour of the slope $\alpha$ vis-a-vis the other free parameters. A large slope implies a predominance of smaller energies, while a small slope implies predominance of larger energies.  

To quantify relations in terms of the peak intensity of a nanoflare, \cite{PSModel} defined the average \thirdround{peak} nanoflare radiance ($E_{avg}$) as:
\begin{equation} \label{eq:eng}
    E_{avg} := \left(\frac{1-\alpha}{2-\alpha}\right).\left(\frac{y_{max}^{2-\alpha}-y_{min}^{2-\alpha}}{y_{max}^{1-\alpha}-y_{min}^{1-\alpha}}.\right)
\end{equation} 

$E_{avg}$ is a measure of the average peak nanoflare radiance value for a given time series. Using Eq.~\ref{eq:eng}, we estimate the values of $E_{avg}$ for each pixel, and study its relationship with flaring frequency ($p_f$) as well as flare duration ($\tau$), through scatter plots between these parameters.

\begin{figure}[h!]
    \plotone{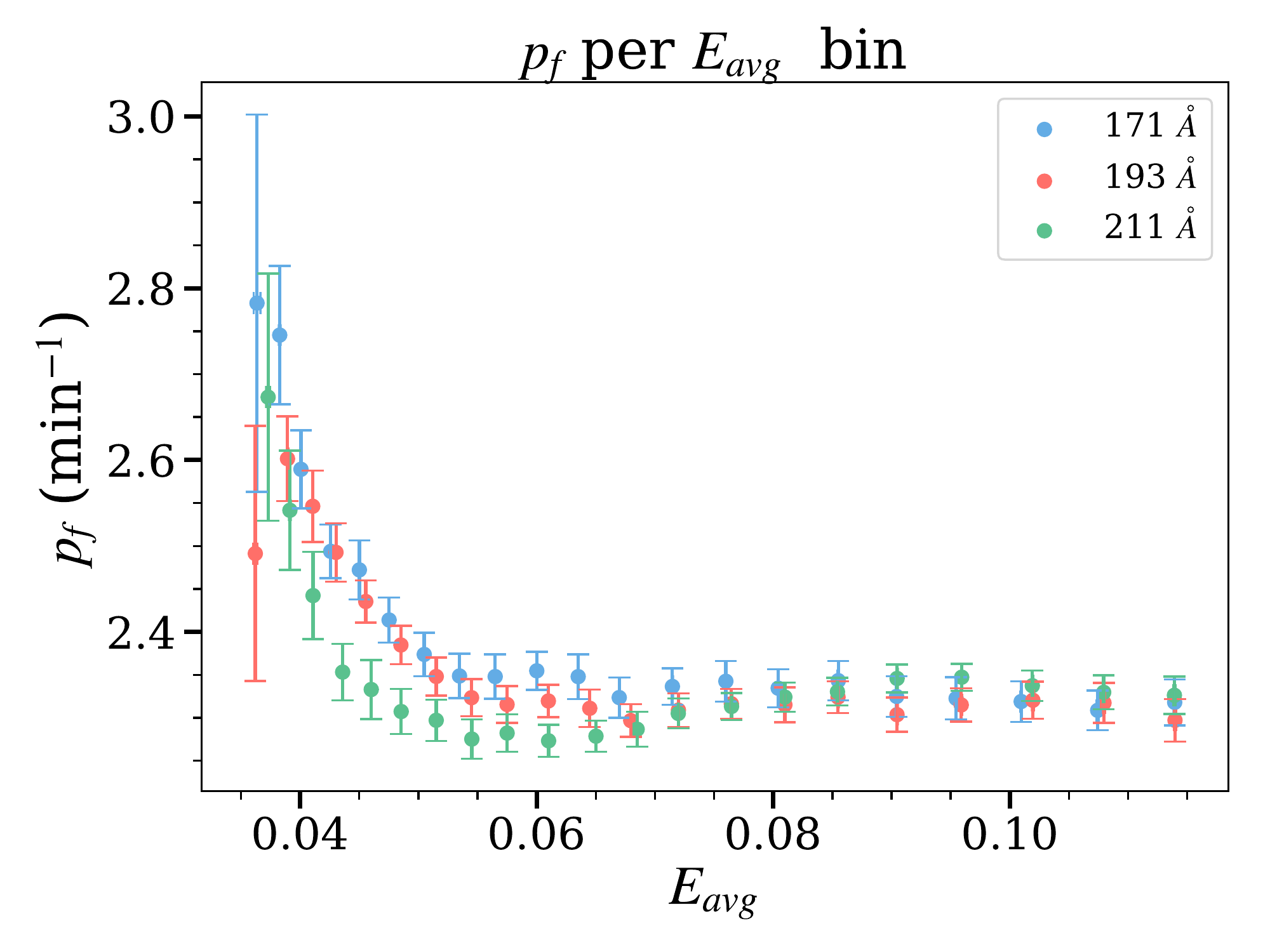}
    \plotone{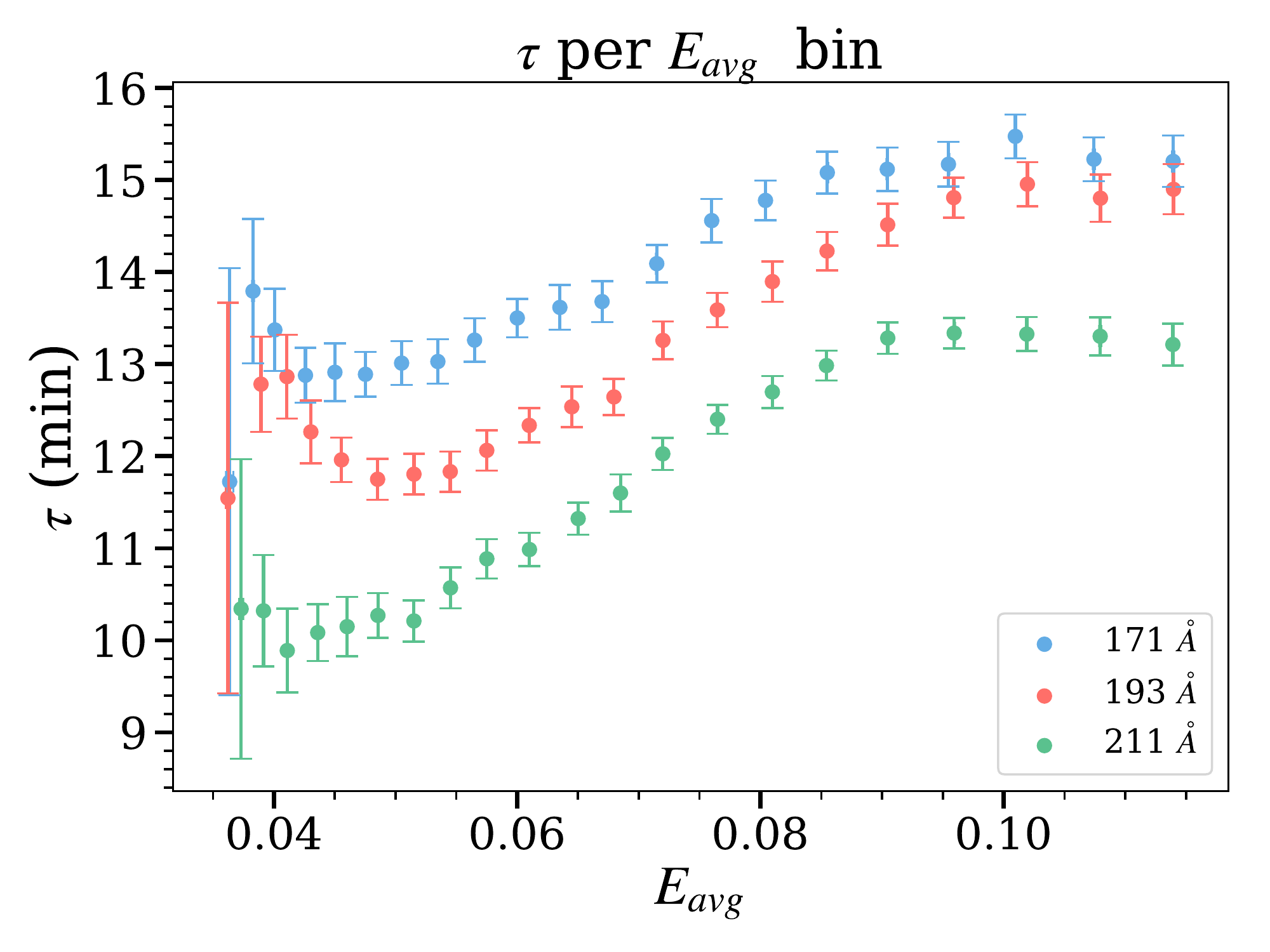}
    \caption{Inferred dependence of free parameters (binned) on $E_{avg}$, over different passbands. Blue colour represents 171~{\AA}, red for 193~{\AA} and green for 211~{\AA}. Left: $p_f$ is plotted per $E_{avg}$ bin; Right: $\tau$ is plotted per $E_{avg}$ bin. The error bars indicate $3\sigma$.} \label{fig:param}
\end{figure}

In Fig.~\ref{fig:param}, we plot $p_f$ (left panel) and $\tau$ (right panel) as a function of $E_{avg}$ for all three wavelengths. Note that we have averaged the free parameters within a constant bin of $E_{avg}$, and our errorbars are $3\sigma$ standard errors.

The plots reveal a slight tendency of decreasing p$_f$ as function of E$_{avg}$, albeit there is sharp decline in the beginning. However, $\tau$ monotonically increases with increasing E$_{avg}$ till about E$_{avg} = 0.085$ and shows saturation thereafter. Moreover, the plot further suggests a systematic lowering of flaring duration, being largest for coolest passband (171~{\AA}).

\section{Summary, Discussion and Conclusion}\label{sec:discussion}

QS coronal region provides a wealth of data to narrow down and understand the underlying heating processes. Assuming the underlying heating mechanism to be impulsive, coronal light curves may be approximated using empirical statistical models to infer physics-inspired parameters. To this end, we construct an inversion model of the \texttt{PSM} using CNN and validate it using a separate test set. We perform the inversion on two coronal datasets and infer the free parameters for the three AIA pass-bands viz. 171~{\AA}, 193~{\AA} and 211~{\AA}.

We find that the light curves inverted using the CNN \vishal{and observed light curves are statistically in excellent statistical agreement}, considering the CDF and KDEs (see Fig.~\ref{fig:replc_ds1} \& \ref{fig:replc_ds2}). \vishal{Note that we are not concerned with a point-to-point match at each timestep between the simulation and the observation. A change in the seed of the random number generator is enough to shift the location of individual events and their amplitudes in simulation. However, this does not cause any changes in the statistical properties of the light curve, which is what we are primarily interested in. The Global wavelet power shows the simulations and observations to have peaks at similar frequencies, validating that the simulation and observation both have enhanced power in similar frequencies}. The quality of approximation may be understood by the Epistemic uncertainty of our CNN, obtained by the application of Dropout to perturb the model. 

We find the distribution of all parameters to be similar for light curves from all three passbands. The flaring frequency lies within the \vishal{range of 1 to 4 events per minute, with a peak at $\sim 2.3$} events per minute for all three passbands. Similarly, the flaring duration has a range of values between 5 and 30 minutes. However, unlike the flaring frequency, the peak of the distribution of flaring duration shows temperature dependence, being highest ($\sim$16 min) for coolest 171~{\AA} channel, $\sim$14 min for 193~{AA} and lowest ($\sim$12 min) for the hottest 211~{AA} channel. 

The power-law index $\alpha$ has a range of values between 1 and 8. The distribution of $\alpha$ peaks at 2.3 for all three passbands of AIA viz. 171, 193, and 211~{\AA}. This finding strongly suggests that nanoflare heating is indeed a viable source of energy in the quiet corona \cite[see, e.g.,][]{hudson1991solar}. We also find the fraction of light curves giving $\alpha\ge2$ progressively reduces from cooler to hotter passbands. \vishal{We further note that} there are a significant number of pixels where $\alpha<2$. \vishal{This is suggestive that low energy events may not be dominant everywhere. \thirdround{However, note that the viability of these low energy events also relies on the flaring frequency $p_f$, requiring a full analysis of the energetics.}  }

Our finding further suggests that there is a change of dynamics for pixels with $\alpha<1.8$ and $\alpha\ge1.8$ (from Fig.~\ref{fig:pftau_alpha}). In the former case, \vishal{$p_f\tau$} is nearly constant, while in the latter case, it reduces with increasing $\alpha$. This may also be observed in Fig.~\ref{fig:param}. Note that $\alpha=2$ corresponds nearly to $E_{avg}=0.08$ and the right tail of $E_{avg}$ represents $\alpha<2$. Here, a definite increase of $p_f$ with decreasing $E_{avg}$ is seen, which is interpreted as excitation increasing with decreasing $E_{avg}$ (and thus increasing $\alpha$). However, we also find that $\tau$ reduces with reducing $E_{avg}$. Thus, the increase in damping counters that in excitation in $\alpha\ge2$ regime, causing a reduction in the ratio in Fig.~\ref{fig:pftau_alpha}. Thus, the increase is excitation is essentially nullified by the increase in damping.

\vishal{We note that our  inferred $p_f$ distribution peaks at $\sim2-3$ events per minute, while \cite{safari1} obtained a mean $p_f$ of $\sim$ 0.33 events per min. This may be explained by the better temporal cadence and spatial resolution of our data, whereby our simulation captures much smaller transient events. However, note that our inferred $p_f$ corresponds to an average waiting time of $\sim30$ sec, which is much smaller than the waiting times  of $\sim230$~s suggested from simple geometric arguments \citep{klimchuk2015}.}

\vishal{There is a discrepancy in the $\tau_d$ (decay time) and $\alpha$ derived here with those obtained by \citet{safari1}. The $\tau_d$ being about a factor of six smaller in our case ($\sim60$ min in \citet{safari1} to $\sim$10 min in our case) and $\alpha$ peaking near $2.3$ in our case, compared to a mean $\alpha$ of 2.6 in \citet{safari1}.} This discrepancy may be attributed to the fact that the data used in this study are at much higher spatial and temporal resolution than \cite{safari1}. Therefore, we must have captured smaller events with much shorter timescales (as also alluded to by \cite{safari1}). Similarly, the discrepancy in power-law index $\alpha$ may be attributed to the high spatial resolution data used in the present work. Moreover, in our simulation, we are sampling flares within a larger energy range with larger $y_{max}/y_{min}$ than those by of \cite{safari1}). \vishal{However, note that our obtained cooling time scales ($\sim600$ sec) are indeed of the order of cooling time scales in the corona obtained by \cite{klimchuk2015}($\sim1000$sec). }

Next, we find that the $p_f$ decreases with $E_{avg}$ (see left panel in Fig.~\ref{fig:param}). The decrease of $p_f$ with $E_{avg}$ is interpreted as a decrease in peak energy released per flare with increasing frequency. Thus, we can either have intermittent, high energy events or sustained low energy events. This variation of $p_f$ with $E_{avg}$ \vishal{is similar to the observation} of the relation obtained between peak flare flux and waiting times by~\citet{hudsonwaitingtime}~\citep[see also][]{SarkarNanditaLorentzBuildup}. \vishal{Furthermore, the inverse relation between $p_f$ and $E_{avg}$, (or a direct relation between the waiting time and succeeding nanoflare energy) was a necessary ingredient needed to reproduce observed EM distribution with temperature in ARs by \cite{Cargill_2014}}. Since the rise time is given as a fraction of the decay time in this setup, we do not distinguish between pre-flaring and post-flaring times. Thus, this may point to the presence of a reservoir of energy that may be exhausted by frequent, small-energy events or intermittent, large-energy events. However, we emphasize that the change in $p_f$ is tiny (2-3 events per min, as can be seen from Fig.~\ref{fig:param}) when compared to the total time scale of these events (10-15 min across all passbands). Thus, we must take this interpretation with caution.

The time scale $\tau$ is seen to increase with $E_{avg}$ (see right panel in Fig.~\ref{fig:param}). This shows an increase in flare time scale corresponds to an increase in average flare energy. This weak correlation is similar to the weak correlation observed between peak flux and flare time scale by~\citet[][see Fig. 3]{Time_peak_relation}. \vishal{However, we may seek to explain this relation qualitatively as below:}

For an iso-thermal, optically thin plasma, the observed intensity is directly proportional to electron number density~\citep{ODwyer}, i.e $DN \propto n_e^2$. From \cite{CargillNanoflare}, we find that:
\begin{equation}
\tau_c \propto n_e:  \text{Conductive cooling dominated plasma}
\end{equation}
and 
\begin{equation}
\tau_r \propto n_e^{-1};\text{Radiative cooling dominated plasma}
\end{equation}

\noindent where $\tau_c$ and $\tau_r$ are conductive and radiative cooling times, respectively.

Combining the equations of timescale and intensity, we obtain:
\begin{equation}
\tau_r \propto \frac{1}{\sqrt{DN}}, \tau_c \propto \sqrt{DN},
\end{equation}

\vishal{Thus, for a conduction cooling dominated plasma, the timescale $\tau$ should increase with the emitted intensity, while for a radiative cooling dominated plasma, we expect the opposite.} Our results show a direct relation between $\tau$ and the peak flare intensity $E_{avg}$ and \vishal{qualitatively} suggest that in such events, conduction losses are dominant over radiative losses, assuming a constant flaring frequency. This is similar to the results obtained by \citet{rajhans2021,guptasarkartripathi,srividya} \vishal{for tiny transient brightenings}. 

\vishal{Finally, the flaring time scales are seen to be largest in the coolest passband, and are seen to decrease from the cooler to hotter passbands. This indicates the decreasing dominance of conduction loss over radiative loss (but the conduction loss still dominates)}, as would be the case for cooling loops \citep[see e.g.,][]{klimchuk2006solving,viall2012,Tripathi2009,GuptaTripathiMason}. We emphasize that this is true under the assumption of constant flaring frequency since only $\sim$ 2-3 events are happening per minute, whereas our total (rise+fall) time in consideration $\sim$ 15 min.

It is important to emphasize that the  \texttt{PSM} is very well suited for explaining the observed light curves from the quiet Sun region obtained at much higher temporal and spatial resolution than was initially studied by \cite{PSModel}. We note that although we have improved the inference by quantifying the uncertainties, the \texttt{PSM} may further be developed by incorporating the plasma filling factor and effective area in the forward model, as has also been suggested in the original paper~\citep{PSModel}. \thirdround{Note that in the \texttt{PSM}, the radiance distribution is considered to be same as energy distribution. However, note that there may be a number of further smaller events which may or may not produce detectable signatures in the intensity images.  It is also possible that many events produce signature in one passband, and not in another.  Therefore the distribution reported using the radiance is just a lower limit of the total number of events.  Hence, incorporation of filling factors in the \texttt{PSM} is an important next step in improving the model. }

Moreover, note that we have fixed the $y_{max}/y_{min}$, limiting the energy range considered for the simulation. Therefore, it would be prudent to keep this ratio as a variable quantity. We have tried to develop an inversion model with varying $y_{min}$ over three orders of magnitude -- however, the inversion model did not give adequate performance (the $R^2$ for $\alpha$ was $\sim0.88$ on the testing set). Thus, introducing the radiance ratio as a free parameter creates a large degeneracy in parameter space~\citep[already been hinted to by][]{PSModel}, which cannot be solved by our neural network or the optimization scheme. Hence, further modeling and training strategies are required to disentangle such highly degenerate parameter spaces. Furthermore, it has been reported that the \thirdround{distribution of flare waiting time (inverse of flaring frequency) follow a time-dependent Poisson process, which gives rise to a power law-tail of waiting time distributions}~\citep{wheatland_waitingtime}. Thus, the assumption of a constant flaring frequency (which translates to exponential waiting time distribution) may also need to be revised in the \texttt{PSM}. 

\vishal{It is important to note that a lot of the magnetic field strength is below the 10 Gauss limit (see Figs.~\ref{fig:replc_ds1} \& \ref{fig:replc_ds2}. However, the variability in the magnetic field strength suggests that the corona is extremely dynamic even in the quiet regions. Hence, for any study to characterize the heating of quiet corona, we need magnetic field measurement with much higher sensitivity and at greater resolution.}

\vishal{Finally, note that while our current analysis has been focused on the QS corona, \cite{safari2} have applied \texttt{PSM} to AR corona successfully, although using data at a very low spatial and temporal resolution. Thus, we expect our results to also extend to AR corona. Furthermore,} it would be interesting to apply our inversion model to chromospheric observations~\citep[see also][]{Jess_2014_chromosphere}. This may be performed on spectral lines with Interface Region Imaging Spectrograph~\citep[IRIS]{iris}, or using chromospheric observations in Near-UltraViolet (NUV) from the Solar Ultraviolet Imaging Telescope \citep[SUIT;][]{suit2,suit1}, onboard the Aditya-L1 mission~\citep{seetha2017aditya} of Indian Space Research Organization (ISRO).

\acknowledgements \vishal{We sincerely thank the referee for constructive comments and suggestions.} This work is partly supported by the Max-Planck Partner Group of Max-Planck Institute for Solar System Research, G\"ottingen, Germany. We acknowledge the use of data from AIA and HMI. AIA is an instrument onboard SDO, a mission for NASA's Living With a Star program. We would like to sincerely thank Dr. A. Pauluhn for sharing the IDL codes of the \texttt{PSM}, which was rewritten and parallelized in python. The authors also thank Prof. Ranjiv Misra of IUCAA for discussion and comments on degeneracy in error surface and scale-free processes.

\software{The analysis was primarily performed in python, using the Tensorflow~\citep{abadi2016tensorflow} package for machine learning -- data download alone was done using Solarsoft~\citep{solarsoft}. Computational tools used were Numpy~\citep{numpy1,numpy2,numpy_nature}, Scipy~\citep{scipy} and Astropy~\citep{astropy1,astropy2}. Parallelization was performed using Multiprocessing~\citep{multiprocessing}, and plotting using Matplotlib~\citep{matplotlib} and Seaborn~\citep{waskom2020seaborn}. Most of the code has been written in Jupyter environment~\citep{jupyter}. \newref{The codebase is under expansion for future work, and shall be made available on request.} }

\bibliography{sample63}{}
\bibliographystyle{aasjournal}

\end{document}